\documentclass[12pt]{iopart}
\usepackage{color}
\usepackage{graphics}
\usepackage{epsf}
\usepackage{graphicx,epsfig}
\def\cs2{c_{s}^{2}}

 \def\al{\alpha}
 \def\b{\beta}
 \def\ga{\gamma}
 \def\de{\delta}
 
 \def\ep{\varepsilon}

 \def\df{\delta\phi}
 \def\p{\partial}

\def\DB{\delta B}
 \def\be   {\begin{equation}}   \def\ee   {\end{equation}}
 \def\ba   {\begin{array}}      \def\ea   {\end{array}}
 \def\bea  {\begin{eqnarray}}   \def\eea  {\end{eqnarray}}
 \def\bean {\begin{eqnarray*}}  \def\eean {\end{eqnarray*}}
\begin{document}

\title{Anisotropic bispectrum of curvature perturbations from primordial non-Abelian vector fields}
\vspace{0.8cm}

\author{Nicola Bartolo$^{1}$, Emanuela Dimastrogiovanni$^{1}$, \\ Sabino Matarrese$^{1}$ and 
Antonio Riotto$^{2}$}
\vspace{0.4cm}
\address{$^1$ Dipartimento di Fisica ``G. Galilei'', Universit\`a degii Studi di
Padova, and INFN Sezione di Padova, via Marzolo 8, I-35131 Padova, Italy.} 

\address{$^2$ CERN, Theory Division, CH-1211 Geneva 23, Switzerland, and INFN Sezione di Padova, via Marzolo 8, I-35131 Padova, Italy.}
\eads{\mailto{nicola.bartolo@pd.infn.it}, \mailto{dimastro@pd.infn.it}, 
\mailto{sabino.matarrese@pd.infn.it} and \mailto{riotto@mail.cern.ch}}

\date{\today}
\vspace{0.8cm}
\begin{abstract}
We consider a primordial $SU(2)$ vector multiplet during inflation in models where quantum fluctuations of vector fields are involved in producing the curvature perturbation. Recently, a lot of attention has been paid to models populated by vector fields, given the interesting possibility of generating some level of statistical anisotropy in the cosmological perturbations. The scenario we propose is strongly motivated by the fact that, for non-Abelian gauge fields, self-interactions are responsible for generating extra terms in the cosmological correlation functions, which are naturally absent in the Abelian case. We compute these extra contributions to the bispectrum of the curvature perturbation, using the $\delta N$ formula and the Schwinger-Keldysh formalism. The primordial violation of rotational invariance (due to the introduction of the $SU(2)$ gauge multiplet) leaves its imprint on the correlation functions introducing, as expected, some degree of statistical anisotropy in our results.\\ 
We calculate the non-Gaussianity parameter $f_{NL}$, proving that the new contributions derived from gauge bosons self-interactions can be important, and 
in some cases the dominat ones. We study the shape of the bispectrum and we find that it turns out to peak in the local configuration, 
with an amplitude that is modulated by the preferred directions that break statistical isotropy. \\
\\
\\
DFPD 09-A-12 / CERN-PH-TH/2009-097

\end{abstract}

\maketitle

\section{Introduction}

The standard cosmological model is based on the idea of inflation as driven by a single slowly-rolling scalar field whose 
primordial quantum fluctuations are responsible for both CMB perturbations and the large scale structures of the universe~\cite{lrreview}. 
Current observations indicate that the universe on large scales is homogeneous and isotropic. The degree of deviation from this
smooth background is provided by the temperature fluctuations of the CMB which are of the order of $10^{-5}$ and are almost 
scale-independent~\cite{smoot92,bennett96,gorski96,wmap3,wmap5}. The standard inflationary scenario predicts the observed 
power spectrum of cosmic fluctuations. On the other hand, there is still a lot of room for alternative scenarios without ruining the 
current agreement with observations. Alternatives include, for example, multifield 
models~\cite{Linde:1985yf,Kofman:1985zx,Polarski:1994rz,GarciaBellido:1995qq,Mukhanov:1997fw,Langlois:1999dw,Gordon:2000hv}, 
the curvaton scenario~\cite{Mollerach,Enqvist:2001zp,Lyth:2001nq,Lyth:2002my,Moroi:2001ct,Bartolo:2003jx} 
and theories with non-canonical Lagrangians as k-inflation~\cite{Garriga:1999vw,ArmendarizPicon:1999rj}, 
DBI inflation~\cite{Ali,Chen:2006nt} or ghost inflation~\cite{ghostinfl}. So far, observations have not provided us with the precision 
measurements which would allow to discriminate among all these different models, but new projects, such as the Planck~\cite{planck} 
satellite which was just launched, can potentially reach the levels of precision necessary for this purpose. One important feature of 
the CMB anisotropies to be decoded is the degree of non-Gaussianity~\cite{review,prop}. A random field is defined as Gaussian if 
all the information is contained in its two-point correlation function. In single-field, slow-roll inflation the level of non-Gaussianity is 
very small since the connected parts of higher order correlation functions are proportional to powers of the slow-roll parameters
~\cite{Acqua,Maldacena:2002vr,Seery:2006vu}. If a larger non-Gaussianity will be detected in the future, this could then rule out the minimal 
model. Recently another alternative to the standard inflationary scenario has 
attracted some attention where primordial vector fields play a 
non-negligible role during inflation 
~\cite{Dimopoulos:2006ms,Ack,Golovnev:2008cf, 
Dimopoulos:2008rf,Yokoyama:2008xw,Dimopoulos:2008yv, 
instability1,Karciauskas:2008bc,instability2,Golovnev:2009ks} 
. An interesting possibility 
is that these extra fields might be responsible for generating an observable level 
of non-Gaussianity~\cite{Dimopoulos:2008yv, 
Karciauskas:2008bc,Yokoyama:2008xw}. 
This is not the only motivation to consider such 
models. First of all vector fields are present in particle physics models 
and that suggests that one has to include them in a theory of the early 
universe. This can be done without necessarily affecting current 
observations, for example keeping their contribution to the energy density 
of the universe way below the one from the inflaton. 
However, a peculiarity of the vector fields, which the scalar fields do 
not possess, is to violate rotational invariance as some observations seem 
to suggest. Violation of rotational invariance would 
affect the correlation functions introducing some degree of statistical 
anisotropy. The power spectrum is constrained in this sense if a single preferred direction of space is involved~\cite{Groeneboom:2008fz}. This is something to keep in mind for example when constucting models of vector inflation. The isotropic background expansion can be preserved to a good approximation in different ways, such as by keeping the energy density of the vector subdominant w.r.t. the one of the inflaton, or by considering the existence of many randomly oriented vectors~\cite{Golovnev:2008cf}. Nevertheless, the presence of the vector fields introduces statistical anisotropy in the power spectrum and in the bispectrum as well~\cite{Dimopoulos:2008yv}. On the other hand, no observational constraints are currently available for the two-point function if more than one preferred direction is involved. Similarly, for higher order correlators, no obervational constraints of this kind are available.\\

The effects of a primordial vector field on the power spectrum and bispectrum of the curvature fluctuations $\zeta$ were recently investigated in~\cite{
Dimopoulos:2008yv,Karciauskas:2008bc} using  the $\delta N$ formalism~\cite{deltaN1,deltaN2,deltaN3,deltaN4} both in vector inflation and in the vector curvaton models, 
and also in \cite{Yokoyama:2008xw} for a model of hybrid inflation.\\

Up to now only primordial Abelian vector fields were considered. In this paper  we wish to explore a more realistic model of gauge interactions of an $SU(2)$ 
vector multiplet. The reason for considering a non-Abelian theory relies on the fact that, differently from what happens in the Abelian case, the vector field components are involved in self-interactions as described by cubic and quartic terms in the Lagrangian. Gauge self-interactions naturally produce additional contributions to the bispectrum w.r.t. the Abelian case. The computation of these new contributions represents the main purpose of our work.\\

The paper is organized as follows: in Section $2$ we introduce the $\delta N$ formula for the curvature perturbation and the Lagrangian of our system; in Section $3$ we carry out the calculations of all the terms that do not require the intervention of the Schwinger-Keldysh formula; in Section $4$ we derive the contribution to the bispectrum that arises from gauge bosons self-interactions; in Section $5$ we study the shape of the bispectrum; in Section $6$ we estimate the non-Gaussianity 
parameter $f_{NL}$; in Section $7$ we draw our conclusions. We include four Appendices: A, on background and first order perturbation equations for the gauge multiplet; B, where we derive the expression of the number of e-foldings of inflation and its derivatives in the presence of vector fields; C, which collects lengthy expressions for functions appearing in the final results for the bispectrum; D, where we give some details about the dependence of the bispectrum from the angles between wave and gauge vectors in a sample spatial configuration.

\section{Bispectrum of the curvature perturbation in the $\delta N$ formalism}

In the the $\delta N$ formalism the curvature perturbation $\zeta(\vec{x})$ at a given time $t$ can be interpreted as a geometrical quantity indicating the fluctuations in the local expansion of the universe; in fact, if $N(\vec{x},t^{*},t)$ is the number of e-foldings of expansion evaluated between times $t^{*}$ and $t$, where the initial hypersurface is chosen to be flat and the final one is uniform density, we have 

\bea\label{first}
\zeta(\vec{x},t)= N(\vec{x},t^{*},t)-N(t^{*},t)\equiv \delta N(\vec{x},t).
\eea
The number of e-foldings $N(\vec{x},t^{*},t)$ depends on all the fields and their perturbations on the initial slice. In principle, since the fields are governed by second order differential equations, it should also depend on their first time derivative, but if we assume that slow-roll conditions apply, then the time derivatives will not count as independent quantities. \\
In a theory which includes vector fields, spatial isotropy is not expected to be preserved and therefore the power spectrum and the bispectrum 
can in principle depend both on the length adn direction of the wavevectors $\vec{k}_i$, rather than just on their moduli

\bea
\langle\zeta_{\vec{k}_{{1}}}(t)\zeta_{\vec{k}_{{2}}}(t) \rangle=(2 \pi)^3\delta^{(3)}(\vec{k_{1}}+\vec{k_{2}})P_{\zeta}(\vec{k}),
\\
\label{bisp}
\langle\zeta_{\vec{k}_{{1}}}(t)\zeta_{\vec{k}_{{2}}}(t)\zeta_{\vec{k}_{{3}}}(t) \rangle=(2 \pi)^3\delta^{(3)}(\vec{k_{1}}+\vec{k_{2}}+\vec{k_{3}})B_{\zeta}(\vec{k_{1}},\vec{k_{2}},\vec{k_{3}}).
\eea
In the next section, we are going to spell out the $\delta N$ expansion in terms of the fields of our theory; later we will replace it into Eq.(\ref{bisp}) and calculate the bispectrum to tree-level.

\subsection{Specializing to our theory}

Our theory includes a scalar field $\phi$ playing the role of the inflaton and an $SU(2)$ gauge multiplet $A_{\mu}^{a}$ (a=1,2,3) non-minimally coupled to gravity. The metric of an FRW flat spacetime $ds^{2}=dt^{2}-a^{2}\de_{ij}dx^{i}dx^{j}$ is employed and held unperturbed. The action is

\be\label{ac}\fl
S=\int d^{4}x \sqrt{-g}\left[\frac{m_{P}^{2}R}{2}-\frac{1}{4}g^{\mu\al}g^{\nu\b}\sum_{a=1,2,3}F_{\mu\nu}^{a}F_{\al\b}^{a}-\frac{M^{2}}{2}g^{\mu\nu}\sum_{a=1,2,3}B_{\mu}^{a}B_{\nu}^{a}+L_{\phi}\right],
\ee
where $L_{\phi}$ is the scalar field Lagrangian and

\bea
F_{\mu\nu}^{a}=\p_{\mu}B^{a}_{\nu}-\p_{\nu}B^{a}_{\mu}+g_{c}\ep^{abc}B^{b}_{\mu}B^{c}_{\nu}
\eea
(we name $g_{c}$ the gauge coupling). Notice that $B^{a}_{\mu}$ are the comoving fields, the physical fields being given by $A_{\mu}^{a}=(B_{0}^{a},(1/a)B_{i}^{a})$. The coupling between the gauge bosons and gravity is ``hidden" in the effective mass $M$: by definition $M^{2}= m_{0}^{2}+\xi R$, where $m_{0}$ is the gauge bosons mass and $\xi$ is a numerical factor; using the slow-roll approximation, the Ricci scalar is 
$R=-6\Big[\frac{\ddot{a}}{a}+{\Big(\frac{\dot{a}}{a})}^{2}\Big]\simeq -12H^2$.\\ 
The quantum fluctuations for the vector fields can be expanded in terms of creation and annihilation operators (this is always correct if we are dealing with free gauge boson, their background equations of motion being of Klein-Gordon type)
\bea
\fl
\delta A_{i}^{a}(\vec{x},\eta)=\int \frac{d^{3}q}{(2\pi)^{3}}e^{i\vec{q}\cdot\vec{x}}\sum_{\lambda=L,R,l}\Big[e^{\lambda}_{i}(\hat{q})a_{\vec{q}}^{a,\lambda}\delta A_{\lambda}^{a}(q,\eta)+e^{*\lambda}_{i}(-\hat{q})\left(a_{-\vec{q}}^{a,\lambda}\right)^{\dagger}\delta A_{\lambda}^{*a}(q,\eta)\Big],
\eea
where
\be
\Big[a_{\vec{q}}^{a,\lambda},\left(a_{\vec{q}^{'}}^{a^{'},\lambda^{'}}\right)^{\dagger}\Big]=(2\pi)^{3}\delta_{\lambda,\lambda^{'}}\delta_{a,a^{'}}\delta^{(3)}(\vec{q}-\vec{q}^{'}),
\ee
$d \eta=dt/a(t)$ is the conformal time and 
$\lambda$ labels left, right and longitudinal polarization states.\\
The perturbation of the number of e-folds on large scales is determined by the perturbations of the fields at the initial time and the complete 
$\delta N$ formula then reads

\bea\fl
\zeta(\vec{x},t)= N_{\phi}\df+N^{\mu}_{a}\delta A_{\mu}^{a}+\frac{1}{2}N_{\phi\phi}\left(\df\right)^2+\frac{1}{2}N^{\mu\nu}_{ab}\delta A_{\mu}^{a}\delta A_{\nu}^{b}+N_{\phi a}^{\mu}\df\delta A_{\mu}^{a}+...
\eea
where

\bea\label{NNN}
N_{\phi}\equiv \left(\frac{\p N}{\p \phi}\right)_{t^{*}},\,\,\,\,\,
N^{\mu}_{a}\equiv\left(\frac{\p N}{\p A^{a}_{\mu}}\right)_{t^{*}},\,\,\,\,\,N_{\phi a}^{\mu}\equiv\left(\frac{\p^{2} N}{\p \phi\p A^{a}_{\mu}}\right)_{t^{*}}
\eea
and so on. $N$ is the number of e-foldings between the time $t^{*}$ on the initial hypersurface and the final time $\eta$ at which the bispectrum of the $\zeta$ is being evaluated. A convenient choice, that we will follow for the rest of the paper, is to take $\eta^{*}$ to be close to the time when a given perturbation mode of wavenumber k crosses out the horizon. The coefficients in Eq.(\ref{NNN}) will be calculated in Appendix B. Notice that the e-folding numbers and their derivatives are defined w.r.t. the physical fields, whereas the action (\ref{ac}) is written in terms of the comoving fields $B_{\mu}^{a}$.\\
Plugging the $\delta N$ expansion into Eq.(\ref{bisp}), we get an infinite series of terms. Retaining the tree-level ones (i.e. the terms that are formally equivalent to the product of two power spectra), we are left with 

\bea\label{tri}\fl
\langle\zeta_{\vec{k}_{{1}}}(t)\zeta_{\vec{k}_{{2}}}(t)\zeta_{\vec{k}_{{3}}}(t)\rangle ^{tree}&=&\frac{3}{2}N_{\phi}^{2}N_{\phi\phi}\int d^3 x \langle\df_{\vec{k}_{1}}\df_{\vec{k}_{2}} \int d^{3} q_{1}d^{3} q_{2}e^{-i\vec{x}(\vec{q}_{1}+\vec{q}_{2})}\df_{\vec{q}_{1}}\df_{\vec{q}_{2}}\rangle_{t^{*}}\nonumber\\\fl&+&\frac{3}{2}N_{a}^{i}N^{j}_{b}N^{kl}_{cd}\langle\delta A^{a}_{i,\vec{k}_{1}}\delta A^{b}_{j,\vec{k}_{2}} \int d^{3} q_{1}d^{3} q_{2}e^{-i\vec{x}(\vec{q}_{1}+\vec{q}_{2})}\delta A^{c}_{k,\vec{q}_{1}}\delta A^{d}_{l,\vec{q}_{2}} \rangle_{t^{*}}\nonumber\\\fl&+&6N_{\phi}N_{a}^{i}N_{\phi b}^{j}\langle \df_{\vec{k}_{1}}\delta A^{a}_{i,\vec{k}_{2}} \int d^{3} q_{1}d^{3} q_{2}e^{-i\vec{x}(\vec{q}_{1}+\vec{q_{2}})}\df_{\vec{q}_{1}}\delta A^{b}_{j,\vec{q}_{2}} \rangle_{t^{*}}\nonumber\\\fl&+&N_{\phi}^{2}N_{\phi}\langle \df_{\vec{k}_{1}}\df_{\vec{k}_{2}}\df_{\vec{k}_{3}}\rangle_{t^{*}}\nonumber\\\fl&+&N_{a}^{i}N_{b}^{j}N_{c}^{k}\langle\delta A^{a}_{i,\vec{k}_{1}}\delta A^{b}_{j,\vec{k}_{2}} \delta A^{c}_{k,\vec{k}_{3}}\rangle_{t^{*}}.
\eea
In the previous expansion we have not included terms that would require some kind of coupling between scalar and vector fields, i.e. terms that are proportional to $N_{\phi}N_{a}^{i}N_{b}^{j}$, $N_{\phi}^{2}N_{\phi a}^{i}$, $N_{\phi}N_{a}^{i}N_{bc}^{jk}$, $N_{\phi}^{2}N_{a}^{i}$, $N_{\phi a}^{i}N_{bc}^{jk}$ and $N_{\phi}N_{\phi\phi}N_{a}^{i}$. Also, the terms that are proportional to $N_{0}^{a}\equiv(\p N)/(\p A^{a}_{0})$ do not appear in the previous equation since the $A_{0}^{a}$ fields are set equal to zero. The reason for this will be explained in details in Appendix A (see discussion after Eq~(\ref{backzero})); here we anticipate that, with a zero expectation value for the temporal components of the background gauge fields and with some upper bounds on the $SU(2)$ coupling constant of the theory, it is possible to obtain a slow-roll evolution for the $A_{i}^{a}$ fields during inflation.\\
It is useful to notice that the terms in the first four lines of (\ref{tri}) are also present in the Abelian case, while the one in the last line is strictly 
non-Abelian: it provides a non-zero contribution only thanks to gauge bosons self-interactions.

\section{Power spectrum and the `Abelian' terms of the bispectrum}

It is useful to write the expression for the power spectrum of the curvature perturbation in the presence of a vector multiplet. Let us define
\bea
P_{\pm}^{ab}\equiv (1/2)(P_{R}^{ab}\pm P_{L}^{ab}),
\eea 
where 
\bea
P_{R}^{ab}&\equiv& \delta_{ab}\delta A_{R}^{a}(k,t^{*})\delta A_{R}^{b*}(k,t^{*}),\\
P_{L}^{ab}&\equiv& \delta_{ab}\delta A_{L}^{a}(k,t^{*})\delta A_{L}^{b*}(k,t^{*}),\\
P_{long}^{ab}&\equiv& \delta_{ab}\delta A_{long}^{a}(k,t^{*})\delta A_{long}^{b*}(k,t^{*})
\eea
($\delta A$ indicating the eigenfunctions for the gauge fields). The power spectrum of $\zeta$ can then be written as

\bea\label{power-zeta}
P_{\zeta}(\vec{k})&=&N^{2}_{\phi}P_{\phi}(k)+N_{a}^{i}N_{b}^{j}\left[\delta_{ij}P^{ab}_{+}+\hat{k}_{i}\hat{k}_{j}\left(P^{ab}_{long}-P^{ab}_{+}\right)+i\epsilon_{ijk}\hat{k}_{k}P^{ab}_{-}\right]\nonumber\\&=&P^{iso}(k)\left[1+g^{ab}\left(\hat{k}\cdot\vec{N}_{a}\right)\left(\hat{k}\cdot\vec{N}_{b}\right)+is^{ab}\hat{k}\cdot\left(\vec{N}_{a}\times\vec{N}_{b}\right)\right],
\eea
where
\bea
g^{ab}\equiv \frac{P^{ab}_{long}-P^{ab}_{+}}{N^{2}_{\phi}P_{\phi}+\left(\vec{N}_{c}\cdot\vec{N}_{d}\right)P^{cd}_{+}},\\
s^{ab}\equiv \frac{P^{ab}_{-}}{N^{2}_{\phi}P_{\phi}+\left(\vec{N}_{c}\cdot\vec{N}_{d}\right)P^{cd}_{+}}
\eea
and we have factorized the isotropic part of the power spectrum 

\bea\label{piso}
P^{iso}(k)\equiv N^{2}_{\phi}P_{\phi}(k)+\left(\vec{N}_{c}\cdot\vec{N}_{d}\right)P^{cd}_{+}.
\eea
In Eq.~\ref{power-zeta} $g^{ab}$ and $s^{ab}$ represent the amplitude of statistical anisotropy, along the preferred directions specified by the vectors $\vec{N}_a
\equiv N^i_a$ and $\vec{N}_b$. 
Eq.(\ref{piso}) can also be written in the form

\bea
P^{iso}(k)=N^{2}_{\phi}P_{\phi}\left[1+\beta_{cd}\frac{P^{cd}_{+}}{P_{\phi}}\right],
\eea
introducing the parameter

\bea
\beta_{cd}\equiv \frac{\vec{N}_{c}\cdot\vec{N}_{d}}{N^{2}_{\phi}}.
\eea

Let us now move to the bispectrum. The first three lines of Eq.(\ref{tri}) can be easily evaluated 

\bea\label{class1}\fl
B_{\zeta}(\vec{k_{1}},\vec{k_{2}},\vec{k_{3}})^{tree}&\supset&\frac{1}{2}N_{\phi}^{2}N_{\phi\phi}\left[P_{\df}({k}_{1})P_{\df}({k}_{2})+perms.\right]\nonumber\\\fl&+&\frac{1}{2}N_{a}^{i}N_{b}^{j}N_{cd}^{kl}\left[\Pi^{ac}_{ik}(\vec{k}_{1})\Pi^{bd}_{jl}(\vec{k}_{2})+perms.\right]
\nonumber\\\fl&+&\frac{1}{2}N_{\phi}N_{a}^{i}N_{\phi b}^{j}\left[P_{\df}({k}_{1})\Pi^{ab}_{ij}(\vec{k}_{2})+perms.\right],
\eea
where we have defined 

\be\label{res}
\Pi_{ij}^{ab}(\vec{k}) \equiv T^{even}_{ij}(\vec{k})P_{+}^{ab}+i T^{odd}_{ij}(\vec{k})P_{-}^{ab}+T^{long}_{ij}(\vec{k})P_{long}^{ab}.
\ee
In the previous equations, $T^{even}_{ij}$, $T^{odd}_{ij}$ and $T^{long}_{ij}$ are defined as in~\cite{Dimopoulos:2008yv}

\bea
T^{even}_{ij}(\vec{k}) &\equiv& e_{i}^{L}(\hat{k})e_{j}^{*L}(\hat{k})+e_{i}^{R}(\hat{k})e_{j}^{*R}(\hat{k}),\\
T^{odd}_{ij}(\vec{k}) &\equiv& i \Big[e_{i}^{L}(\hat{k})e_{j}^{*L}(\hat{k})-e_{i}^{R}(\hat{k})e_{j}^{*R}(\hat{k})\Big],\\
T^{long}_{ij}(\vec{k}) &\equiv& e_{i}^{l}(\hat{k})e_{j}^{*l}(\hat{k}),
\eea
with $e^{L}(\hat{k})\equiv\frac{1}{\sqrt{2}} (\cos\theta\cos\phi-i\sin\phi,\cos\theta\sin\phi+i\cos\phi,-\sin\theta)$, $e^{R}(\hat{k})=e^{*L}(\hat{k})$ and $e^{l}(\hat{k})=\hat{k}=(\sin\theta\cos\phi,\sin\theta\sin\phi,\cos\theta)$, so

\bea
T^{even}_{ij}(\vec{k})&=&\delta_{ij}-\hat{k}_{i}\hat{k}_{j},\\
T^{odd}_{ij}(\vec{k})&=&\epsilon_{ijk}\hat{k}_{k},\\
T^{long}_{ij}(\vec{k})&=&\hat{k}_{i}\hat{k}_{j}.
\eea
Notice that if gauge indices are suppressed in Eq.(\ref{class1}), the Abelian expression is recovered (see~\cite{Dimopoulos:2008yv}).\\
It is convenient to rewrite the expression in the second line of Eq.(\ref{class1}) as follows

\bea\fl
N_{a}^{i}N_{b}^{j}N_{cd}^{kl}\Pi^{ac}_{ik}(\vec{k}_{1})\Pi^{bd}_{jl}(\vec{k}_{2})=\left(N_{a}^{i}\Pi_{ik}^{ac}\right)N_{cd}^{kl}\left(N_{b}^{j}\Pi^{bd}_{jl}\right)=M^{c}_{k} N^{kl}_{cd} M^{d}_{l},
\eea 
where

\bea\label{nonzero2}\fl
\vec{M}^{c}(\vec{k})\equiv P^{ac}_{+}(k)\left[\vec{N}_{a}+p^{ac}(k)\hat{k}\left(\hat{k}\cdot\vec{N}_{a}\right)+iq^{ac}(k)\hat{k}\times\vec{N}_{a}\right] 
\eea
and we define
\bea
p^{ac}(k)&\equiv& \frac{P^{ac}_{long}-P^{ac}_{+}}{P^{ac}_{+}},\\
q^{ac}(k) &\equiv& \frac{P^{ac}_{-}}{P^{ac}_{+}}. 
\eea

Similarly, the third line in Eq.(\ref{class1}) becomes

\bea
N_{\phi}N_{a}^{i}N_{\phi b}^{j}P_{\phi}(k_{1})\Pi^{ab}_{ij}(\vec{k}_{2})=N_{\phi}N_{\phi a}^{i}P_{\phi}(k_{1})M^{a}_{i}(k_{2}).
\eea

The contribution from the fourth line in Eq.(\ref{tri}) can be calculated using the Schwinger-Keldysh formalism~\cite{in-in1,in-in2,in-in3}. It is a well-known result 
tha the sum of  this last term and the one from the first line of Eq.(\ref{tri}) is proportional to the slow-roll parameters~\cite{Acqua,Maldacena:2002vr}. \\
It is important to notice that, in models where $P^{ab}_{long}=P^{ab}_{+}$ and $P_{-}^{ab}=0$, the coefficients $g^{ab}$, $s^{ab}$, $p^{ab}$ and $q^{ab}$ are all equal to zero and therefore both the power spectrum and the Abelian part of the bispectrum of
$\zeta$ become isotropic. In general $P_{-}^{ab}=0$, unless there is some mechanism producing parity violation for $A_i$. The condition $P_{ab}^{long}=P^{ab}_{+}$ is more subtle. It could be realized if the longitudinal and the transverse parts of the vector fields evolved in the same way \cite{Dimopoulos:2008yv}.

\section{Calculation of the `non-Abelian' terms of the bispectrum}

The last line of Eq.(\ref{tri}) also requires the Schwinger-Keldysh formula

\be
\langle\Omega|\Theta(t)|\Omega\rangle=\left\langle 0\left|\left[\bar{T}\left(e^{i {\int}^{t}_{0}H_{I}(t')dt'}\right)\right]\Theta_{I}(t)\left[T \left(e^{-i {\int}^{t}_{0}H_{I}(t')dt'}\right)\right]\right|0\right\rangle,
\ee  
where $\Theta(t)$ is a field operator, $|\Omega\rangle$ represents the vacuum of the  interaction theory, $T$ and $\bar{T}$ are time-ordering and anti-time-ordering operators. All the fields are in the interaction picture, as the subscript $I$ indicates. \\Our next step will then be to write explicitly the action for the gauge bosons and look at the detailed expression of their interaction Hamiltonian. Using the definition of $F_{\mu\nu}^{a}$, the action for the gauge fields becomes

\bea\label{ACTION}\fl
S_{A_{\mu}}&=&\int d^{4}x \sqrt{-g}\Big[-\frac{1}{2}g^{\mu\al}g^{\nu\b}\Big(\p_{\mu}B_{\nu}^{a}\p_{\al}B_{\b}^{a}-\p_{\mu}B_{\nu}^{a}\p_{\b}B_{\al}^{a}\Big)\\\fl
&-&g_{c}\ep^{abc}g^{\mu\al}g^{\nu\b}\Big(\p_{\mu}B_{\nu}^{a}\Big)B_{\al}^{b}B^{c}_{\b}-\frac{1}{4}g_{c}^{2}\ep^{abc}\ep^{ade}g^{\mu\al}g^{\nu\b}B_{\mu}^{b}B_{\nu}^{c}B_{\al}^{d}B_{\b}^{e}-\frac{M^{2}}{2}g^{\mu\nu}B_{\mu}^{a}B_{\nu}^{a}\Big].\nonumber
\eea

The interaction Hamiltonian to third order in the gauge field perturbations is thus made up of two contributions coming respectively from the third and fourth order interations in the lagrangian 

\bea\label{HamiltonianINT}
H_{int}=g_{c}\ep^{abc}\Big(\p_{\mu}\de B^{a}_{\nu}\Big)\de B^{b\mu}\de B^{c \nu}+g_{c}^2 \ep^{eab}\ep^{ecd}B^{a}_{\mu}\de B^{b}_{\nu}\de B^{c\mu}\de B^{d\nu}.
\eea
The correction to the bispectrum~(\ref{tri}) due to these interactions has the form

\bea\label{ris}
\fl
B_{\zeta}(\vec{k_{1}}\vec{k_{2}}\vec{k_{3}}) \supset N_{A_{i}^{a}}N_{A_{j}^{b}}N_{A_{k}^{c}}\int d \eta a^{4}(\eta) \tilde{\Pi}_{im}(\vec{k_{1}})\tilde{\Pi}^{l}_{j}(\vec{k_{2}})\tilde{\Pi}^{m}_{k}(\vec{k_{3}})\Big[g_{c}\ep^{abc}k_{1l}+g_{c}^2\ep^{eda}\ep^{ebc}B^{d}_{l}\Big]\nonumber\\+perms.+c.c.
\eea
where 

\be\label{ress}
\tilde{\Pi}_{ij}^{ab}(\vec{k}) \equiv T^{even}_{ij}(\vec{k})\tilde{P}_{+}^{ab}+i T^{odd}_{ij}(\vec{k})\tilde{P}_{-}^{ab}+T^{long}_{ij}(\vec{k})\tilde{P}_{long}^{ab}
\ee
and $\tilde{P}_{\pm}\equiv (1/2)(\tilde{P}_{R}\pm \tilde{P}_{L})$, $\tilde{P}_{R}$ being defined as the product of the eigenfunctions $\delta B_{R}(k,\eta^{*})$ and $\delta B_{R}^{*}(k,\eta)$, and so on for $\tilde{P}_{L}$ and $\tilde{P}_{long}$. Notice that $\eta$ is the time at which the three point function for $\zeta$ is being evaluated. Also, we dropped the gauge indices in $\tilde{P}$, since the perturbations have the same time and momentum dependence for the different gauge fields. This last point can be discussed looking at Eq.~(\ref{perturbationsEQ}) for the perturbations of $\delta B^{a}_{i}$: the equations for the free fields (which are the ones we need to compute the power spectrum and the bispectrum) are obtained from (\ref{perturbationsEQ}) setting $g_{c}=0$, i.e. suppressing the interaction terms, and thus we 
get the same equation for all the components of the gauge multiplet.

\subsection{Bispectrum from third-order interactions}

Let us begin with the bispectrum contribution from Eq.(\ref{ris}) at the lowest order in the coupling $g_{c}$
  
\bea\label{zeta}
\langle \zeta_{\vec{k_{1}}}\zeta_{\vec{k_{2}}}\zeta_{\vec{k_{3}}}\rangle_{\eta} &\supseteq&N^{i}_{a^{'}}N^{j}_{b^{'}}N^{k}_{c^{'}}\langle \delta A^{a^{'}}_{i}(\vec{k_{1}})\delta A^{b^{'}}_{j}(\vec{k_{2}})\delta A^{c^{'}}_{k}(\vec{k_{3}})\rangle_{*}\nonumber\\&=&N^{i}_{a^{'}}N^{j}_{b^{'}}N^{k}_{c^{'}}\frac{\delta^{(3)}(\vec{k_{1}}+\vec{k_{2}}+\vec{k_{3}})}{a^{3}(\eta^{*})}\ep^{abc}\delta_{aa^{'}}\delta_{bb^{'}}\delta_{cc^{'}}\nonumber\\&\times&\int_{-\infty}^{\eta^{*}}d\eta^{'} a^{4}(\eta^{'})\tilde{\Pi}^{aa^{'}}_{il}\tilde{\Pi}^{bb^{'}}_{jm}\tilde{\Pi}^{cc^{'}}_{kn}g^{sm}g^{ln}i k_{s}^{(1)}g_{c}+perms.+c.c
\eea
where $g^{ij}$ is the unperturbed background metric. The factor $i k_{s}^{(1)}$ originates from the spatial derivative in the interaction term considered, whereas $a^{4}$ come from $\sqrt{-g}$ together with having moved from cosmic to conformal time $dt=ad\eta$. In order to write Eq.(\ref{zeta}), we used the third order interaction Hamiltonian

\bea\label{Hint}\fl
H^{(3)}_{int}=g_{c}\ep^{abc}\left[g^{ik}g^{jl}\left(\p_{i}\DB_{j}^{a}\right)\DB^{b}_{k}\DB_{l}^{c}+g^{ij}\left(\p_{i}\DB_{0}^{a}\right)\DB^{b}_{j}\DB_{0}^{c}+g^{ij}\left(\p_{0}\DB_{i}^{a}\right)\DB^{b}_{0}\DB_{j}^{c}\right].
\eea 
The calculation of the tree-level contributions due to the last two vertices in Eq.(\ref{Hint}) will not be carried out since they contain the time components $\DB_{0}^{a}$, which turn out to be affected by instabilities. In fact, $\delta B_{0}$ is a non-dynamical mode that, for the free-field case, can be expressed in terms of the longitudinal component as follows [31] 

\bea
\delta B_{0}=-i\frac{\p_{t}\left(\vec{k}\cdot\delta\vec{B}\right)}{k^2+(aM)^2}.
\eea
The longitudinal modes of the theory have negative kinetic energy and are therefore considered unstable. In addition to this, their equations of motion suffer from singularities around the horizon crossing time. However, a regular solution was provided in \cite{Dimopoulos:2008yv}, which we are going to employ when handling these modes (Eq.~(\ref{longitudinal-mode})). \\The factor $1/a^{3}(\eta^{*})$ outside the integral accounts for the fact that the action is written in terms of the comoving fields $B_{i}^{a}$ whereas the bispectrum of $\zeta$ is written in terms of the bispectrum of the physical fields $A_{i}^{a}$. \\
In our model, $\tilde{P}_{-}=0$ (i.e. $\tilde{P}_{R}=\tilde{P}_{L}$), therefore the integrand of Eq.(\ref{zeta}) can be written as the sum of products of $\tilde{P}_{+}=\tilde{P}_{R}$ and $\tilde{P}_{long}$ weighted by products of coefficients 
$T^{even}$ and $T^{long}$

\bea\fl
T^{EEE}_{ijk}&=& k_{1}\Big[\delta_{ik}\hat{k}_{j1}-\hat{k}_{i1}\hat{k}_{j1}\hat{k}_{k1}-\hat{k}_{i3}\hat{k}_{j1}\hat{k}_{k3}-\delta_{ik}(\hat{k}_{1}\cdot \hat{k}_{2})\hat{k}_{j2}+(\hat{k}_{1}\cdot\hat{k}_{2})\Big(\hat{k}_{i3}\hat{k}_{j2}\hat{k}_{k3}\nonumber\\\fl&+&\hat{k}_{i1}\hat{k}_{j2}\hat{k}_{k1}\Big)+(\hat{k}_{1}\cdot\hat{k}_{3})\hat{k}_{i1}\hat{k}_{j1}\hat{k}_{k3}-(\hat{k_{1}}\cdot\hat{k_{2}})(\hat{k}_{1}\cdot\hat{k}_{3})\hat{k}_{i1}\hat{k}_{j2}\hat{k}_{k3}\Big]\\\fl
T^{EEl}_{ijk}&=&k_{1}\Big[\hat{k}_{i3}\hat{k}_{j1}\hat{k}_{k3}-\hat{k}_{i3}\hat{k}_{j2}\hat{k}_{k3}(\hat{k}_{1}\cdot \hat{k}_{2})-\hat{k}_{i1}\hat{k}_{j1}\hat{k}_{k3}(\hat{k}_{1} \cdot \hat{k}_{3})\nonumber\\\fl&+&\hat{k}_{i1}\hat{k}_{j2}\hat{k}_{k3}(\hat{k}_{1}\cdot \hat{k}_{2})(\hat{k}_{1}\cdot \hat{k}_{3})\Big] \\\fl
T^{ElE}_{ijk}&=&k_{1}\Big[\delta_{ik}\hat{k}_{j2}(\hat{k}_{1}\cdot \hat{k}_{2})-\hat{k}_{i3}\hat{k}_{j2}\hat{k}_{k3}(\hat{k}_{1}\cdot\hat{k}_{2})-\hat{k}_{i1}\hat{k}_{j2}\hat{k}_{k1}(\hat{k}_{1}\cdot\hat{k}_{2})\nonumber\\\fl&+&\hat{k}_{i1}\hat{k}_{j2}\hat{k}_{k3}(\hat{k}_{1}\cdot\hat{k}_{3})(\hat{k}_{1}\cdot\hat{k}_{2})\Big] \\\fl
T^{lEE}_{ijk}&=&k_{1}\Big[\hat{k}_{i1}\hat{k}_{j1}\hat{k}_{k1}-\hat{k}_{i1}\hat{k}_{j1}\hat{k}_{k3}(\hat{k}_{1}\cdot\hat{k}_{3})-\hat{k}_{i1}\hat{k}_{j2}\hat{k}_{k1}(\hat{k}_{1}\cdot\hat{k}_{2})\nonumber\\\fl&+&\hat{k}_{i1}\hat{k}_{j2}\hat{k}_{k3}(\hat{k}_{1}\cdot\hat{k}_{2})(\hat{k}_{1}\cdot\hat{k}_{3})\Big] \\\fl
T^{llE}_{ijk}&=& k_{1}\Big[\hat{k}_{i1}\hat{k}_{j2}\hat{k}_{k1}(\hat{k}_{1}\cdot\hat{k}_{2})-\hat{k}_{i1}\hat{k}_{j2}\hat{k}_{k3}(\hat{k}_{1}\cdot\hat{k}_{2})(\hat{k}_{1}\cdot\hat{k}_{3})\Big]\\\fl
T^{lEl}_{ijk}&=& k_{1}\Big[\hat{k}_{i1}\hat{k}_{j1}\hat{k}_{k3}(\hat{k}_{1}\cdot\hat{k}_{3})-\hat{k}_{i1}\hat{k}_{j2}\hat{k}_{k3}(\hat{k}_{1}\cdot\hat{k}_{2})(\hat{k}_{1}\cdot\hat{k}_{3})\Big]\\\fl
T^{Ell}_{ijk}&=& k_{1}\Big[\hat{k}_{i3}\hat{k}_{j2}\hat{k}_{k3}(\hat{k}_{1}\cdot\hat{k}_{2})-\hat{k}_{i1}\hat{k}_{j2}\hat{k}_{k3}(\hat{k}_{1}\cdot\hat{k}_{2})(\hat{k}_{1}\cdot\hat{k}_{3})\Big]\\\fl
T^{lll}_{ijk}&=& k_{1}\Big[\hat{k}_{i1}\hat{k}_{j2}\hat{k}_{k3}(\hat{k}_{1}\cdot\hat{k}_{2})(\hat{k}_{1}\cdot\hat{k}_{3})\Big] 
\eea
where where we used the notation $T^{EEl}_{ijk}\equiv T^{even}_{il}T^{even}_{jm}T^{long}_{kn}\delta^{sm}\delta^{ln}k_{s}$ and so on.\\
The next step before carrying out the final integration over time in Eq.(\ref{zeta}) is to sum over all the permutations separately for each one of the eight $T_{ijk}^{\alpha\beta\gamma}$ coefficients. Because of the antisymmetric properties of $\ep^{abc}$, the sums provide results of the form $\ep^{abc}S_{ab}$, where $S_{ab}$ is a symmetric tensor. We thus find out that, independently of the results of time integrals (i.e. independently of the particular form of the wavefunctions for transverse and longitudinal modes), the contribution to the bispectrum from third-order interactions is zero.\\ 
Let us then move to considering the quartic interaction. 

\subsection{Bispectrum from fourth-order interactions}

The contribution from the quartic interaction is

\bea\label{equat1}\fl
\langle\zeta_{\vec{k}_{1}}\zeta_{\vec{k}_{2}}\zeta_{\vec{k}_{3}} \rangle &\supset& (2 \pi)^3 \delta^{3}(\sum_{i=1,2,3}\vec{k}_{i})g_{c}^{2}\frac{1}{H}{\left(\frac{Hx^{*}}{k}\right)}^{3}N_{a'}^{i}N_{b'}^{j}N_{c'}^{k}A^{e}_{l}\epsilon^{aa'c'}\epsilon^{ab'e}\nonumber\\\fl&\times&\sum_{\alpha,\beta,\gamma}\left(\int dx \right)_{\alpha \beta\gamma}T^{\alpha}_{ip}T^{\beta}_{jp}T^{\gamma}_{kl}+perms.+c.c.\nonumber\\\fl&=&(2 \pi)^3 \delta^{3}(\sum_{i=1,2,3}\vec{k}_{i})g_{c}^{2}\frac{1}{H}{\left(Hx^{*}\right)}^{3} \sum_{\alpha\beta\gamma}K_{\alpha\beta\gamma}+perms.
\eea
with $x=-k \eta$. $(Hx^{*})^3$ comes from the $1/a^{3}(\eta^{*})$ factor outside the time integral and the $1/H$ factor comes from having a factor of $a$ inside the time integral, i.e. $\int d\eta a(\eta)=-\int dx/(Hx)$. This factor originates from the background field appearing inside the integral (the interaction term is quartic and can be developed as a product of three perturbations times a background field), i.e. $B^{a}_{l}(\eta)=a(\eta)A^{a}_{l}(\eta)$, where $A$ is the slowly-rolling physical field (and hence taken out of the integral). Notice that, once again, we are setting $A_{0}^{a}=0$, so terms proportional to $N^{0}_{a}$ do not appear in the formulas. The indices $\alpha,\beta,\gamma$ as usual run over the longitudinal and the transverse components. $\left(\int dx \right)_{\alpha \beta\gamma}$ is a shorthand for the integral over the wavefunctions associated with the internal legs times the wavefunctions associated with the external ones. The definition of the symbol $K_{\alpha\beta\gamma}$ can be read from the first two lines in Eq.(\ref{equat1}). It is important to sum over all of the permutations, i.e. over all the possible field contractions, which we are going to list below

\bea\label{a1}
c_{1}^{\alpha\beta\gamma}=T_{ip}^{\alpha}T_{jl}^{\beta}T_{kp}^{\gamma}\delta_{a'b}\delta_{b'c}\delta_{c'd} \\
c_{2}^{\alpha\beta\gamma}=T_{ip}^{\alpha}T_{jp}^{\beta}T_{kl}^{\gamma}\delta_{a'b}\delta_{b'd}\delta_{c'c}\\
c_{3}^{\alpha\beta\gamma}=T_{il}^{\alpha}T_{jp}^{\beta}T_{kp}^{\gamma}\delta_{a'c}\delta_{b'b}\delta_{c'd}\\
c_{4}^{\alpha\beta\gamma}=T_{il}^{\alpha}T_{jp}^{\beta}T_{kp}^{\gamma}\delta_{a'c}\delta_{b'd}\delta_{c'b}\\
c_{5}^{\alpha\beta\gamma}=T_{ip}^{\alpha}T_{jp}^{\beta}T_{kl}^{\gamma}\delta_{a'd}\delta_{b'b}\delta_{c'c}\\\label{a2}
c_{6}^{\alpha\beta\gamma}=T_{ip}^{\alpha}T_{jl}^{\beta}T_{kp}^{\gamma}\delta_{a'd}\delta_{b'c}\delta_{c'b}.
\eea

Let us write the expression of the wavefunctions of the gauge fields. We are going to work in the regime where $M^2=-2H^2$, i.e. we take $m_{0} \ll H$ and $\xi=1/6$. This is the situation where the transverse mode of the $\delta\vec{B}$ field is governed by an equation of motion similar to the one of a light scalar field in unperturbed FRW background (see Eq.(\ref{transverse})). The solution is very well known and, after matching with the vacuum solution at early times, is given by

\be\label{transverse-mode}
\DB^{T}=-\frac{\sqrt{\pi x}}{2\sqrt{k}}\left[J_{3/2}(x)+iJ_{-3/2}(x)\right].
\ee   
where $x\equiv - k \eta$. We have dropped the gauge index since the equation of motion is unique for all of the gauge bosons (see Eq.(\ref{transverse}) and (\ref{longitudinal}) and discussion below Eq.~(\ref{ress})).\\
As to the longitudinal mode, there is an on-going discussion about its instability~\cite{Dimopoulos:2008yv,instability1,instability2,Golovnev:2009ks}; according to some 
authors~\cite{instability1,instability2}, there are two sources of instability: the first one is due to the fact that the equation of motion (\ref{longitudinal}) is singular for a given mode $k$ equal to the effective mass of the vector field; the second one arises from the field behaving like a 'ghost', i.e. having a negative energy. In spite of the divergence affecting the linearized equation of motion, the authors of~\cite{Dimopoulos:2008yv} provided two independent exact solutions to (\ref{longitudinal}). The initial conditions were formulated after rescaling the vector field with the introduction of $\delta\tilde{B}\equiv(a|M|/k)\delta B^{||}$, which allows for the well-known initial conditions $\delta\tilde{B}=e^{-ik\eta}/\sqrt{2k}$ at very early times. We are going to use the solution and the initial condition prescription of~\cite{Dimopoulos:2008yv}, so the longitudinal mode function acquires the following form

\bea\label{longitudinal-mode}
\delta B^{||}=\frac{1}{2\sqrt{k}}\left(x-\frac{2}{x}+2i\right)e^{ix}.
\eea
 
The terms $K_{\alpha\beta\gamma}$ below (permutations are included) then read

\bea\label{Kfirst}\fl
K_{EEE}&=& -\frac{I_{EEE}}{24k^{6}k^{2}_{1}k^{2}_{2}k^{2}_{3}x^{*5}}\left[A_{EEE}+\left(B_{EEE}\cos x^{*}+C_{EEE}\sin x^{*}\right)E_{i} x^{*}\right] \\\fl
K_{lll}&=& -\frac{I_{lll}}{192k^{9}k^{3}_{1}k^{3}_{2}k^{3}_{3}x^{*5}}\left[A_{lll}+\left(B_{lll}\cos x^{*}+C_{lll}\sin x^{*}\right)E_{i}x^{*}\right] \\\fl
K_{llE}&=& \frac{I_{llE}}{96k^{8}k^{3}_{1}k^{3}_{2}k^{2}_{3}x^{*5}}\left[A_{llE}+\left(B_{llE}\cos x^{*}+C_{llE}\sin x^{*}\right)E_{i}x^{*}\right]  \\\label{Klast}\fl
K_{EEl}&=& -\frac{I_{EEl}}{48k^{7}k^{2}_{1}k^{2}_{2}k^{3}_{3}x^{*3}}\left[A_{EEl}+\left(B_{EEl}\cos x^{*}+C_{EEl}\sin x^{*}\right)E_{i}x^{*}\right] 
\eea
where $A$, $B$, $C$ and $D$ are functions (to be provided in Appendix C) of $x^{*}$ and of the momenta $k_{i}\equiv |\vec{k}_{i}|$, $i=1,2,3$ and $E_{i}$ is the exponential-integral function. \\The factors $I_{\alpha\beta\gamma}$ are given by 
\bea\label{anisotropic-coefficients1}\fl
I_{EEE}&\equiv&\ep^{aa'b'}\ep^{ac'e}\Big[6\left(\vec{N}^{a'}\cdot\vec{N}^{c'}\right)\left(\vec{N}^{b'}\cdot\vec{A}^{e}\right)\nonumber\\\fl&+&\left(\vec{N}^{b'}\cdot\vec{A}^{e}\right)\Big[\Big(-2\left(\hat{k}_{3}\cdot\vec{N}^{a'}\right)\left(\hat{k}_{3}\cdot\vec{N}^{c'}\right)-2\left(\hat{k}_{1}\cdot\vec{N}^{a'}\right)\left(\hat{k}_{1}\cdot\vec{N}^{c'}\right)\nonumber\\\fl&+&\left(\hat{k}_{1}\cdot\vec{N}^{a'}\right)\left(\hat{k}_{3}\cdot\vec{N}^{c'}\right)\hat{k}_{1}\cdot\hat{k}_{3}+\left(\hat{k}_{3}\cdot\vec{N}^{a'}\right)\left(\hat{k}_{1}\cdot\vec{N}^{c'}\right)\hat{k}_{1}\cdot\hat{k}_{3}\Big)+(1\rightarrow 2)+(3\rightarrow 2)\Big]\nonumber\\\fl&-&\left[\left(2\left(\vec{N}^{a'}\cdot\vec{N}^{c'}\right)\left(\hat{k}_{2}\cdot\vec{N}^{b'}\right)\left(\hat{k}_{2}\cdot\vec{A}^{e}\right)\right)+(2\rightarrow 1)+(2\rightarrow 3)\right]\nonumber\\\fl&+&\Big[\Big(\hat{k}_{2}\cdot\vec{A}^{e}\Big[2\left(\hat{k}_{3}\cdot\vec{N}^{a'}\right)\left(\hat{k}_{2}\cdot\vec{N}^{b'}\right)\left(\hat{k}_{3}\cdot\vec{N}^{c'}\right)+2\left(\hat{k}_{1}\cdot\vec{N}^{a'}\right)\left(\hat{k}_{2}\cdot\vec{N}^{b'}\right)\left(\hat{k}_{1}\cdot\vec{N}^{c'}\right)\nonumber\\\fl&-&\left(\hat{k}_{1}\cdot\vec{N}^{a'}\right)\left(\hat{k}_{2}\cdot\vec{N}^{b'}\right)\left(\hat{k}_{3}\cdot\vec{N}^{c'}\right)\hat{k}_{1}\cdot\hat{k}_{3}-\left(\hat{k}_{1}\cdot\vec{N}^{a'}\right)\left(\hat{k}_{2}\cdot\vec{N}^{b'}\right)\left(\hat{k}_{3}\cdot\vec{N}^{c'}\right)\hat{k}_{1}\cdot\hat{k}_{3}\Big]\Big)\nonumber\\\fl&+&(2\leftrightarrow 1)+(3\leftrightarrow 2)\Big]\Big] \\\fl
I_{lll}&\equiv&\ep^{aa'b'}\ep^{ac'e}\Big[\Big(\left(\hat{k}_{1}\cdot\vec{N}^{a'}\right)\left(\hat{k}_{3}\cdot\vec{N}^{b'}\right)\left(\hat{k}_{2}\cdot\vec{N}^{c'}\right)\left(\hat{k}_{1}\cdot\hat{k}_{2}\right)\left(\hat{k}_{3}\cdot\vec{A}^{e}\right)\nonumber\\\fl&-&\left(\hat{k}_{3}\cdot\vec{N}^{a'}\right)\left(\hat{k}_{2}\cdot\vec{N}^{b'}\right)\left(\hat{k}_{1}\cdot\vec{N}^{c'}\right)\left(\hat{k}_{1}\cdot\hat{k}_{2}\right)\left(\hat{k}_{3}\cdot\vec{A}^{e}\right)\Big)+(1\leftrightarrow 3)+(2\leftrightarrow 3) \Big]\\\fl
I_{llE}&\equiv& \ep^{aa'b'}\ep^{ac'e}\Big[\left(\vec{N}^{b'}\cdot\vec{A}^{e}\right)\Big(\left(\hat{k}_{1}\cdot\vec{N}^{a'}\right)\left(\hat{k}_{2}\cdot\vec{N}^{c'}\right)+\left(\hat{k}_{2}\cdot\vec{N}^{a'}\right)\left(\hat{k}_{1}\cdot\vec{N}^{c'}\right)\Big)\hat{k}_{1}\cdot\hat{k}_{2}\nonumber\\\fl&+&\left[\Big(2\left(\hat{k}_{2}\cdot\vec{A}^{e}\right)\left(\hat{k}_{1}\cdot\vec{N}^{a'}\right)\left(\hat{k}_{2}\cdot\vec{N}^{b'}\right)\left(\hat{k}_{1}\cdot\vec{N}^{c'}\right)\Big)+(1\leftrightarrow 2)\right]\nonumber\\\fl&-&\Big[\Big(\left(\left(\hat{k}_{1}\cdot\vec{N}^{a'}\right)\left(\hat{k}_{2}\cdot\vec{N}^{c'}\right)+\left(\hat{k}_{2}\cdot\vec{N}^{a'}\right)\left(\hat{k}_{1}\cdot\vec{N}^{c'}\right)\right)\left(\hat{k}_{3}\cdot\vec{N}^{b'}\right)\left(\hat{k}_{3}\cdot\vec{A}^{e}\right)\hat{k}_{1}\cdot\hat{k}_{2}\Big)\nonumber\\\fl&+&(1\leftrightarrow 3)+(2\leftrightarrow 3)\Big]\Big]\\\label{anisotropic-coefficients2}\fl
I_{EEl}&\equiv& \ep^{aa'b'}\ep^{ac'e}\Big[4\left(\vec{N}^{b'}\cdot\vec{A}^{e}\right)\left(\hat{k}_{3}\cdot\vec{N}^{a'}\right)\left(\hat{k}_{3}\cdot\vec{N}^{c'}\right)\nonumber\\\fl&+&\Big[\Big(\left(\hat{k}_{2}\cdot\vec{N}^{b'}\right)\left(\hat{k}_{2}\cdot\vec{A}^{e}\right)\left(\left(\hat{k}_{1}\cdot\vec{N}^{a'}\right)\left(\hat{k}_{3}\cdot\vec{N}^{c'}\right)+\left(\hat{k}_{3}\cdot\vec{N}^{a'}\right)\left(\hat{k}_{1}\cdot\vec{N}^{c'}\right)\right)\hat{k}_{1}\cdot\hat{k}_{3}\Big)\nonumber\\\fl&+&(2\leftrightarrow 1)+(2\leftrightarrow 3)\Big]\nonumber\\\fl&-&\left[\left(2\left(\hat{k}_{2}\cdot\vec{A}^{e}\right)\left(\hat{k}_{2}\cdot\vec{N}^{a'}\right)\left(\hat{k}_{3}\cdot\vec{N}^{b'}\right)\left(\hat{k}_{2}\cdot\vec{N}^{c'}\right)\right)+(1\leftrightarrow 2)+(2\leftrightarrow 3)+(1\leftrightarrow 3)\right]\nonumber\\\fl&-&\left[\left(\left(\vec{N}^{b'}\cdot\vec{A}^{e}\right)\hat{k}_{1}\cdot\hat{k}_{3}\left(\left(\hat{k}_{1}\cdot\vec{N}^{a'}\right)\left(\hat{k}_{3}\cdot\vec{N}^{c'}\right)+\left(\hat{k}_{3}\cdot\vec{N}^{a'}\right)\left(\hat{k}_{1}\cdot\vec{N}^{c'}\right)\right)\right)+(1\leftrightarrow 2)\right] \nonumber\\\fl&+&\left[\left(\vec{N}^{a'}\cdot\vec{N}^{b'}\right)\left(\hat{k}_{3}\cdot\vec{N}^{c'}\right)\left(\hat{k}_{3}\cdot\vec{A}^{e}\right)+\left(\vec{N}^{c'}\cdot\vec{N}^{b'}\right)\left(\hat{k}_{3}\cdot\vec{N}^{a'}\right)\left(\hat{k}_{3}\cdot\vec{A}^{e}\right)\right]\Big]
\eea
where $i \rightarrow j$ means `replace $\hat{k}_{i}$ with $\hat{k}_{j}$', whereas $i \leftrightarrow j$ means `exchange $\hat{k}_{i}$ with $\hat{k}_{j}$'. \\
The terms that were not included in the list above, i.e. $K_{lEE}$, $K_{ElE}$, $K_{Ell}$, $K_{lEl}$, are derived from $K_{EEl}$ and $K_{llE}$ by an appropriate exchange of the momentum labels, for example $K_{lEl}$ and $K_{ElE}$ are obtained respectively from $K_{llE}$ and $K_{EEl}$ by permuting $\hat{k}_{2}$ with $\hat{k}_{3}$, $K_{Ell}$ and $K_{lEE}$ from $K_{llE}$ and $K_{EEl}$ by permuting $\hat{k}_{1}$ with $\hat{k}_{3}$.\\
As one can see through  Eq.~(\ref{Kfirst}) to~(\ref{anisotropic-coefficients2}), the resulting bispectrum is made up of isotropic parts that 
depend only on the moduli of the wavevectors $k_i$ and that are modulated by the coefficients $I_{\alpha \beta \gamma}$. These coefficients 
contain the information about the breaking of rotational invariance, depending on the angles between the wavevectors and the preferred directions $\vec{N}^a$.

Given the fact that the instability of the longitudinal mode of vector perturbations is still considered a controversial issue, a possible alternative to the exact solution provided in~\cite{Dimopoulos:2008yv} is to parametrize the longitudinal mode in terms of the transverse solution as follows 
\bea
\delta B^{||}=n(x)\delta B^{T}
\eea
where $n$ is an unknown function of $x\equiv -k \eta$. The most relevant contribution to the time integrals in Eq.(\ref{equat1}) is due to the region around horizon crossing $x^{*}$ (see for example~\cite{Maldacena:2002vr}), therefore a good approximation when integrating can be to treat $n$ as a constant evaluating it at $x=x^{*}$. The sum over $K_{\alpha\beta\gamma}$ then becomes
\bea\label{approximation-n}\fl
\sum_{\alpha\beta\gamma}K_{\alpha\beta\gamma}=\frac{K_{EEE}}{I_{EEE}}\left[I_{EEE}+n^{2}(x^{*})\left(I_{EEl}+I_{ElE}+I_{lEE}\right)+n^{4}(x^{*})\left(I_{llE}+I_{lEl}+I_{Ell}\right)+n^{6}(x^{*})I_{lll}\right]\nonumber\\
\eea
The accuracy of this approximation has been tested considering again the solutions given in Eq.(\ref{transverse-mode}) and (\ref{longitudinal-mode}), which satisfy the relation $\delta B^{||}=\sqrt{2}\delta B^{T}$ at horizon crossing. If we set $n(x^*)=\sqrt{2}$ in Eq.(\ref{approximation-n}) and compare each one of the terms in round brackets with the corresponding exact time integrals provided in Eq.(\ref{Kfirst}) through (\ref{Klast}), we get a factor of $2$ to $5$ between the exact time integrals and their approximations for the most part of the $(x_{2},x_{3})$ range.\\ 
The particular case $n(x^{*})=1$, corresponding to  logitudinal and transverse modes which evolve in the same way, is worthy of mention: the anisotropic part of Eq.(\ref{equat1}) cancels out leaving an isotropic contribution only

\bea\label{isotropic-bispectrum-res}\fl
\langle\zeta_{\vec{k}_{1}}\zeta_{\vec{k}_{2}}\zeta_{\vec{k}_{3}} \rangle &\supset& (2 \pi)^3 \delta^{3}(\sum_{i=1,2,3}\vec{k}_{i})g_{c}^{2}\frac{1}{H}{\left(\frac{Hx^{*}}{k}\right)}^{3}\epsilon^{aa'b'}\epsilon^{ac'e}\left(\vec{N}^{a'}\cdot \vec{N}^{c'}\right)\left(\vec{N}^{b'}\cdot \vec{A}^{e}\right)\nonumber\\\fl&\times&\frac{K_{EEE}}{I_{EEE}}+perms.
\eea
The way this result is achieved can be easily seen from the expression of the $T^{\alpha}T^{\beta}T^{\gamma}$ products, from which $K_{\alpha\beta\gamma}$ were evaluated. We list these products for one of the permutations

\bea\fl
T_{il}^{E}T_{jp}^{E}T_{kp}^{E}&=& \delta_{il}\delta_{jp}+\delta_{il}\left(-k_{j3}k_{k3}-k_{j2}k_{k2}+k_{j2}k_{k3}k_{23}\right)-k_{1l}\left(\delta_{jk}k_{i1}-k_{133}-k_{122}+k_{123}k_{23}\right)\nonumber\\\fl
T_{il}^{E}T_{jp}^{E}T_{kp}^{l}&=&\delta_{il}\left(k_{j3}k_{k3}-k_{j2}k_{k3}k_{23}\right)-k_{l1}\left(k_{133}-k_{123}k_{23}\right) \nonumber\\\fl
T_{il}^{E}T_{jp}^{l}T_{kp}^{E}&=&\delta_{il}\left(k_{j2}k_{k2}-k_{j2}k_{k3}k_{23}\right)-k_{l1}\left(k_{122}-k_{123}k_{23}\right)\nonumber \\\fl
T_{il}^{l}T_{jp}^{E}T_{kp}^{E}&=&k_{l1}\left(\delta_{jk}k_{i1}-k_{133}-k_{122}+k_{123}k_{23}\right)\nonumber \\\fl
T_{il}^{l}T_{jp}^{l}T_{kp}^{E}&=& k_{l1}\left(k_{122}-k_{123}k_{23}\right)\nonumber\\\fl
T_{il}^{l}T_{jp}^{E}T_{kp}^{l}&=&k_{l1}\left(k_{133}-k_{123}k_{23}\right) \nonumber\\\fl
T_{il}^{E}T_{jp}^{l}T_{kp}^{l}&=& \delta_{il}k_{j2}k_{k3}k_{23}-k_{l1}k_{123}k_{23}\nonumber\\\fl
T_{il}^{l}T_{jp}^{l}T_{kp}^{l}&=& k_{l1}k_{123}k_{23}\nonumber
\eea
where $k_{abc}\equiv k_{ia}k_{jb}k_{kc}$ and $k_{ab}\equiv \vec{k_{a}}\cdot\vec{k_{b}}$ ($a,b,c=1,2,3$ running over the external momenta). Notice that summing over the eight terms in the equations above only leaves $\delta_{il}\delta_{jp}$.

\section{Shape of the bispectrum}

As we saw in Eq.(\ref{equat1}), the bispectrum in the presence of an $SU(2)$ gauge multiplet can be written in terms of isotropic parts, i.e. functions of $x^{*}$ and of the moduli of the external wave vectors $k_i$, weighted by anisotropic coefficients $I_{\alpha\beta\gamma}$ which depend on the angles between the (wave and gauge) vectors. One way of studying the profile of the bispectrum is to analyse the isotropic parts separately in order to understand what is the preferred configuration, for example if it resembles the local or the equilateral shapes~\cite{Babich:2004gb}. Once the answer to that is found, we calculate the anisotropic coefficients in that specific configuration. Let us employ the variables $x_{2}\equiv k_{2}/k_{1}$ and $x_{3}\equiv k_{3}/k_{1}$ as in~\cite{Babich:2004gb}. Plotting the isotropic part, one can notice that for most of the functions $K_{\alpha\beta\gamma}/I_{\alpha\beta\gamma}$, which sum up to provide the bispectrum in (\ref{equat1}), the regions with the highest values are around $x_{2}=1$, $x_{3}=0$ (see plots in Fig.1). The graphs that don't have their maxima in this region, i.e. $r_{llE}$ and $r_{ElE}$, which peak around $x_{2}=x_{3}=1$, and $r_{lEE}$, which peaks around $x_{2}=x_{3}=0.5$, show negligible amplitudes w.r.t. the previous graphs. Therefore we can safely state that the bispectrum peaks in the `local' configuration $k_{2}\sim k_{1}$, $k_{3}\ll k_{1}$. In fact this result can be understood recalling that the contribution to the bispectrum comes from 
the quartic interaction terms in Eq.~(\ref{HamiltonianINT}), which are local in space, and do not involve gradient terms of the fields.   \\

\begin{figure}\centering
 \includegraphics[width=0.4\textwidth]{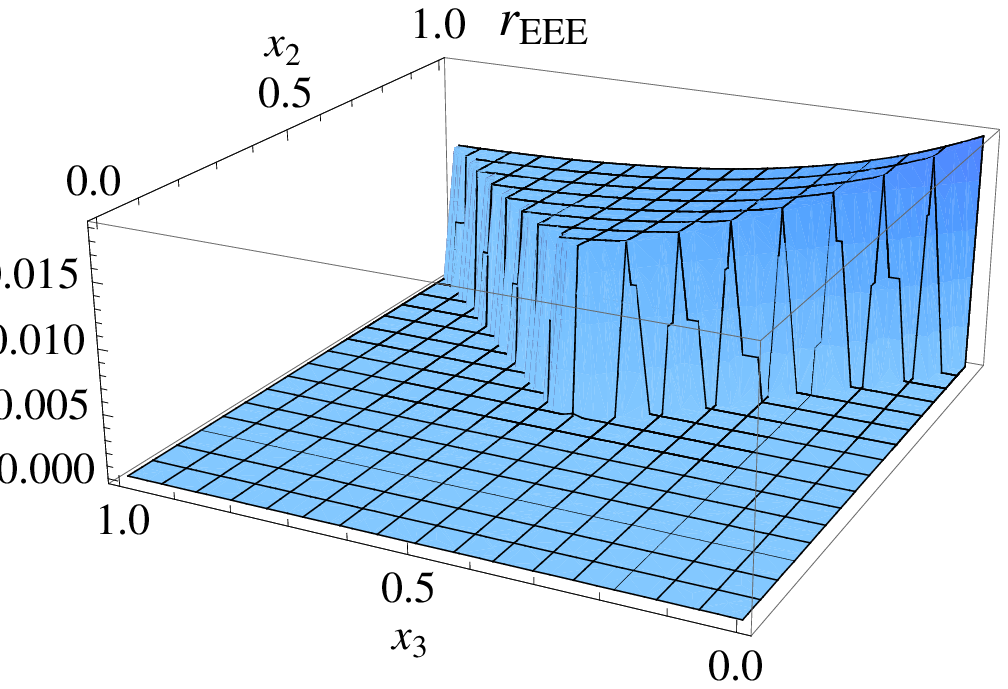}
\hspace{0.1\textwidth}
 \includegraphics[width=0.4\textwidth]{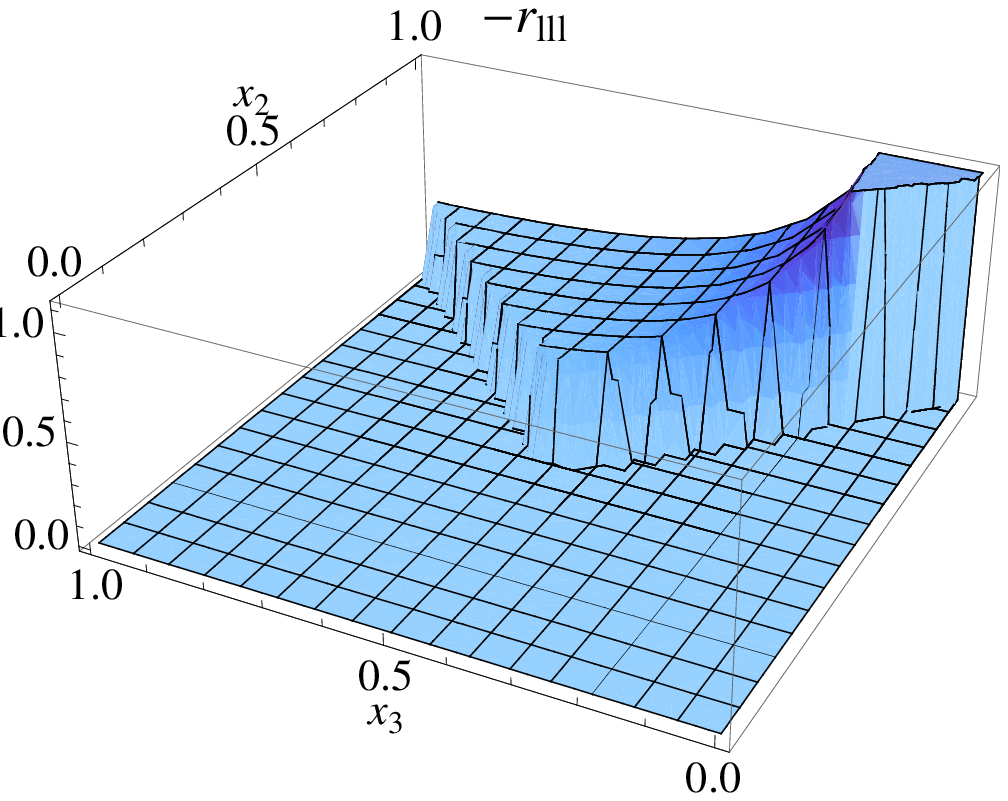}
\vspace{0.02\textwidth}
 \includegraphics[width=0.4\textwidth]{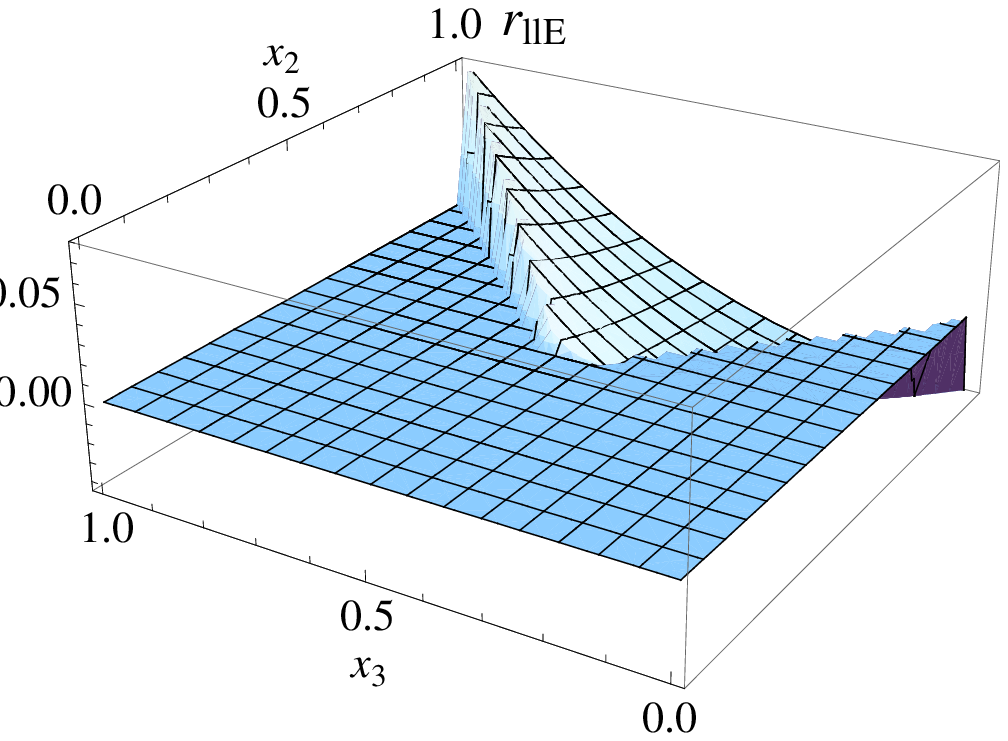}
\hspace{0.1\textwidth}
 \includegraphics[width=0.4\textwidth]{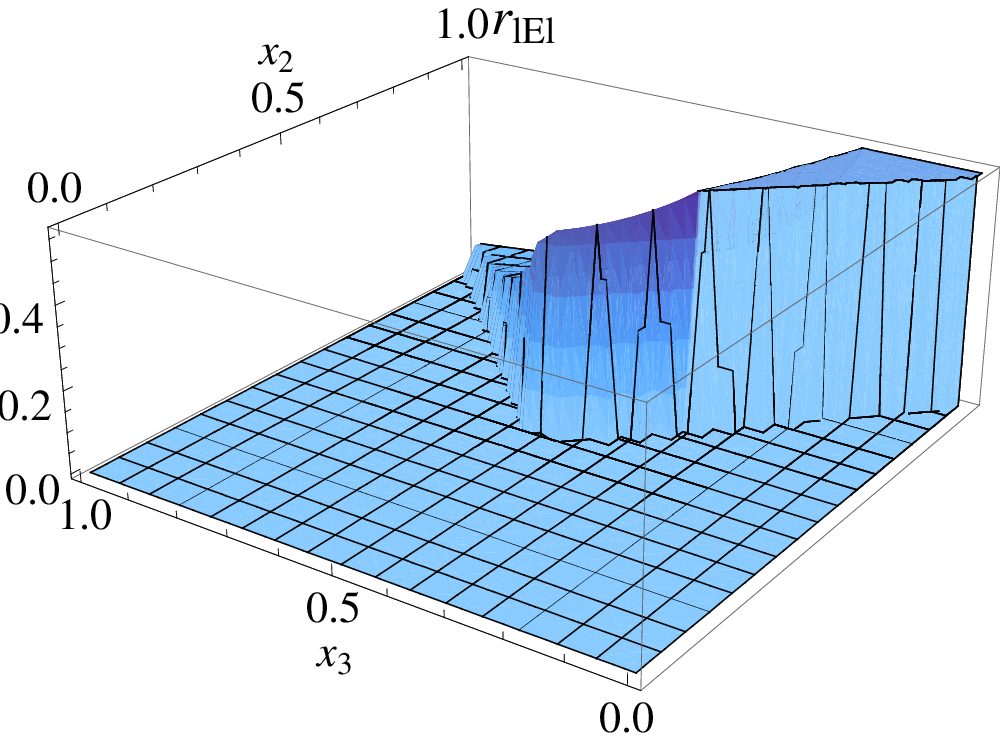}
\vspace{0.02\textwidth}
 \includegraphics[width=0.4\textwidth]{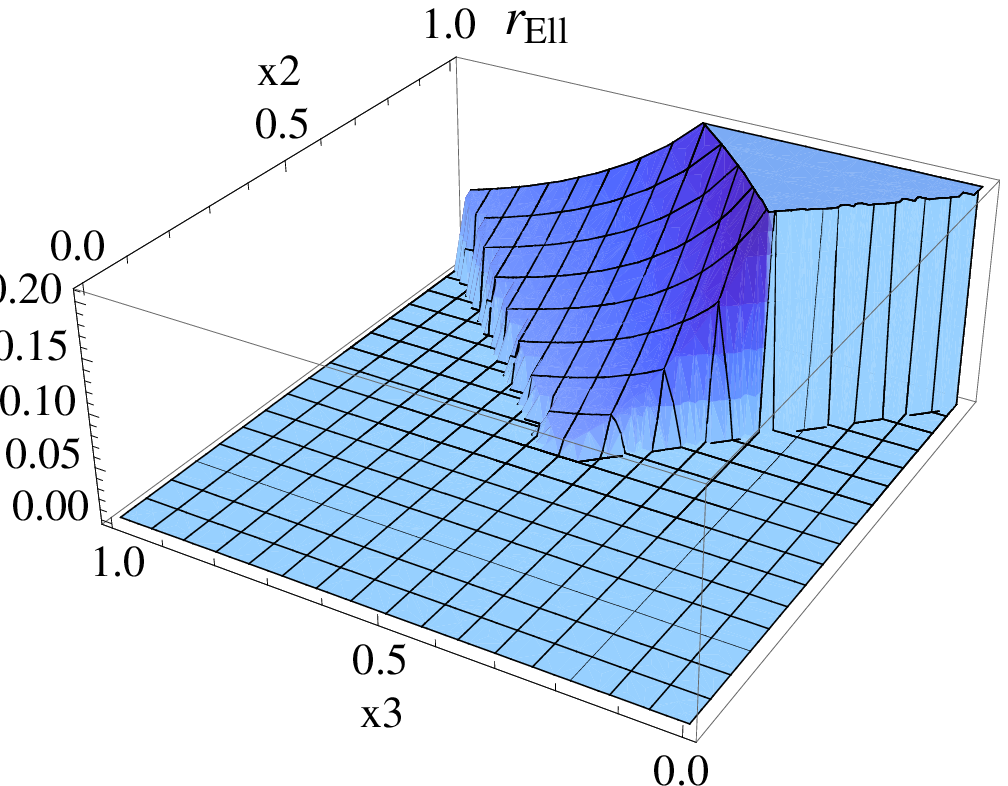}
\hspace{0.1\textwidth}
 \includegraphics[width=0.4\textwidth]{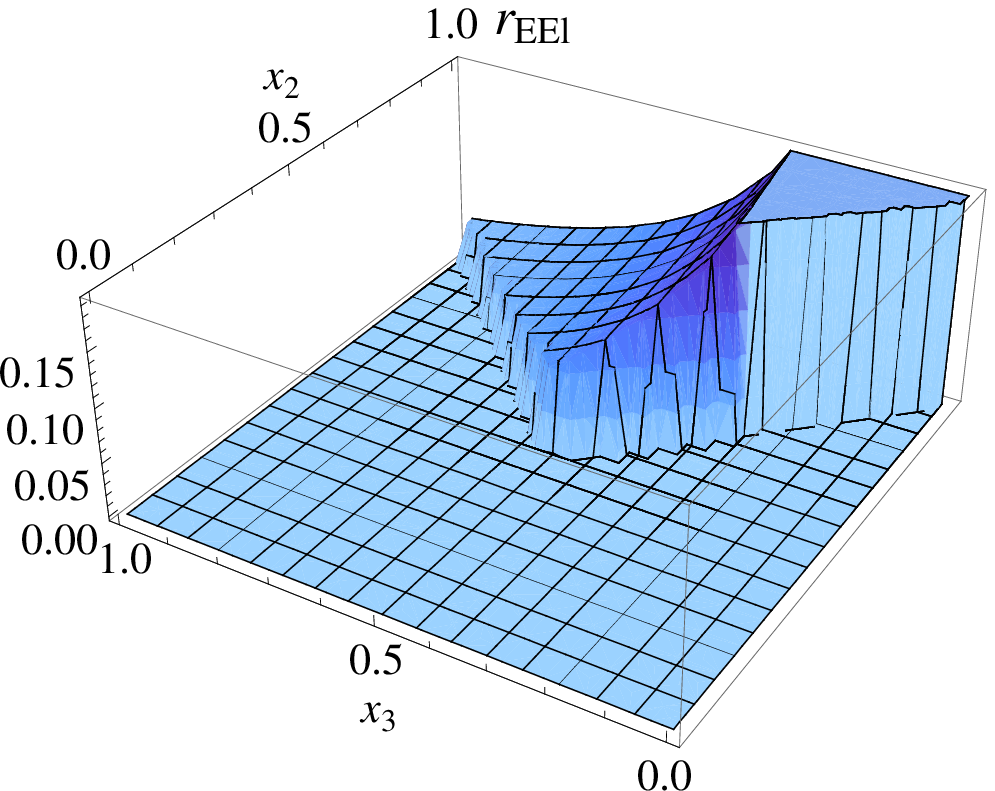}
\vspace{0.02\textwidth}
 \includegraphics[width=0.4\textwidth]{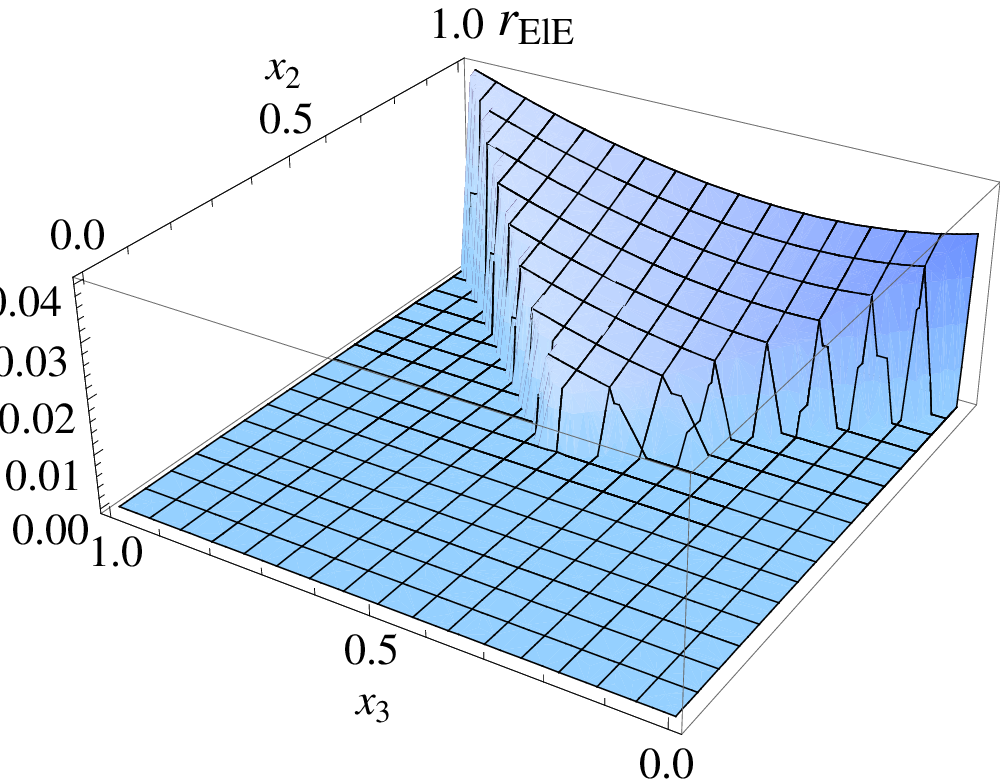}
\hspace{0.1\textwidth}
 \includegraphics[width=0.4\textwidth]{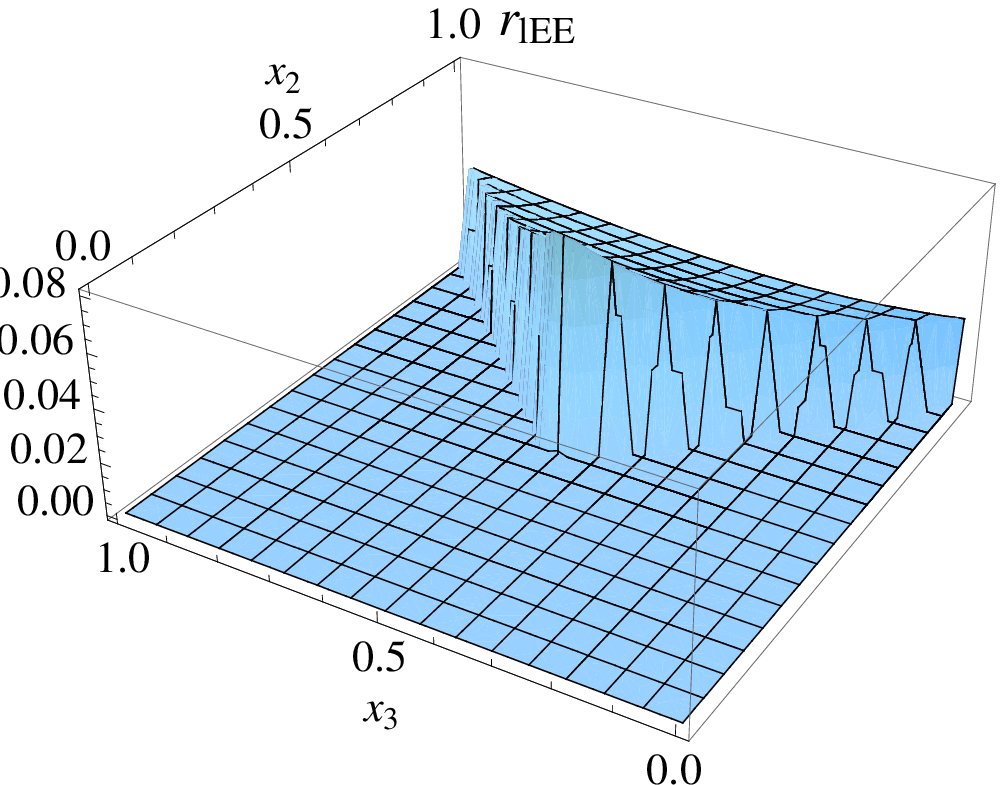}
\caption{ \label{Fig1}Plot of $r_{\alpha\beta\gamma}\equiv \Theta(x_{2}-x_
{3})\Theta(x_{3}-1+x_{2}) x_{2}^{2}x^{2}_{3}R_{\alpha\beta\gamma}(x_{2},x_{3})$,$\qquad$\\
where we define $R_{\alpha\beta\gamma}=k_{1}^{6}(K_{\alpha\beta\gamma}/I_
{\alpha\beta\gamma})$. The Heaviside step functions\\ $\Theta$ help
restricting the plot domain to the region $(x_{2},x_{3})$ that is allowed\\
for the triangle $\vec{k}_{1}+\vec{k}_{2}+\vec{k}_{3}=0$ (in particular, we
set $x_{3}<x_{2}$).We also\\ set $x^{*}=1$.}
\end{figure}

We can now calculate the anisotropic coefficients in the local configuration. We have the freedom to pick up a coordinate system $(\hat{x},\hat{y},\hat{z})$ with $\hat{k}_{1}$ and $\hat{k}_{2}$ pointing along $\hat{z}$ and $\hat{k}_{3}$ along $\hat{x}$. In the general case of three differently oriented vectors $\vec{N}_{a}$, the coefficients $I_{\alpha\beta\gamma}$ depend on a number of angular variables that is bigger than two and, therefore, they are not easy to plot. What we can do is looking at some particular cases with specific orientations of the $\vec{N}_{a}$ w.r.t one another and to the coordinate frame we built from the wave-vectors. \\
Let us consider the simplest situation with the three vectors $\vec{N}_{a}$ aligned along a given direction of space with spherical coordinates $(\theta,\phi)$. It is easy to show that in this situation all of the coefficients $I_{\alpha\beta\gamma}$ are equal to zero, so the bispectrum contribution from (\ref{equat1}) becomes zero as well. Notice that this does not apply either to the power spectrum or to the Abelian part of the bispectrum, which retain a non-zero anisotropic part (see Eqs.(\ref{power-zeta}) and (\ref{nonzero2})) when the vectors $\vec{N}_{a}$ are parallel to each other.\\ 
There are several other situations that can be taken into account, for example one of the $\vec{N}_{a}$ being aligned along or perpendicular to one of the $\hat{k}_{i}$ and the other two generically oriented, two of the $\vec{N}_{a}$ aligned along $\hat{k}_{1}$ and $\hat{k}_{3}$ respectively and the third one generically oriented and so on. Our next step is to provide a plot of the bispectrum complete of its isotropic and anisotropic parts in the favoured configuration (in our case the local one). For convenience while plotting, let us normalize the vector $\vec{N}_{a}$ so that they all have the same length and let us restrict our attention to situations where the total number of angles the $I_{\alpha\beta\gamma}$ coefficients depend on does not exceed two. One possibility is for example given by the following case

\bea\label{instance1}
\vec{N}_{3}=N_{A}(0,0,1)\\\label{instance2}
\vec{N}_{1}=\vec{N}_{2}=N_{A}(\sin\theta\cos\phi,\sin\theta\sin\phi,\cos\theta),
\eea
where, as mentioned above, we are working in the coordinate frame with  $\hat{k}_{3}=\hat{x}$ and $\hat{k}_{1}=\hat{k}_{2}=\hat{z}$ and we took $N_{A}\equiv |\vec{N}^{a}|$ for all $a=1,2,3$. Let us introduce the angle $\delta$ between $\vec{N}_{1,2}$ and $\hat{k}_{3}$. Then from Eq.(\ref{equat1}) we have

\bea
B_{\zeta}\supset g_{c}^{2}\frac{H^2}{k_{1}^{6}}\sum_{\alpha\beta\gamma}I_{\alpha\beta\gamma}(\theta,\delta)R_{\alpha\beta\gamma}(x^{*},x_{2},x_{3}).
\eea
The functions $R_{\alpha\beta\gamma}(x_{2},x_{3})$ are plotted in Fig.1 for $x^{*}=1$ and their analytic expressions are provided in Appendix C. See Appendix D for the expressions of the coefficients $I_{\alpha\beta\gamma}(\theta,\delta)$.\\A plot of the 'non-Abelian' bispectrum normalized to the ratio $(g_{c}^{2}H^2 m^2N_{A}^{4})/(k^{6}_{1}x_{2}^{2}x_{3}^{2})$ is given in Fig.2 for fixed values of $x^{*}$, $x_{2}$ and $x_{3}$ (see Appendix D for its analytic expression).

\begin{figure}\centering
 \includegraphics[width=0.4\textwidth]{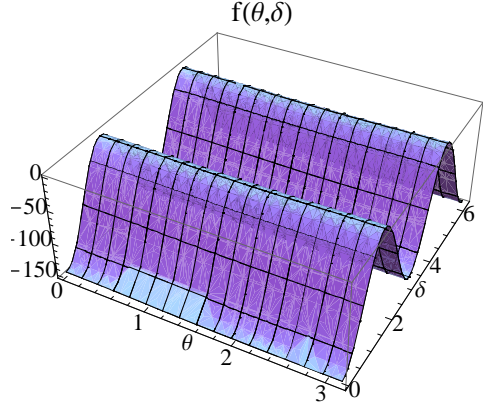}
\caption{ \label{Fig2}Plot of $f(\theta,\delta)\equiv {[(B_{\zeta}(\theta,\delta,x^{*},x_{2},x_{3})x_{2}^{2}x_{3}^{2}k_{1}^{6})/(g_{c}^{2}H^2 m^2 N_{A}^{4})]}_{(x^{*}=1,x_{2}=0.9,x_{3}=0.1)}$. As we can see, the graph dependence from $\theta$ is extremely weak; in fact, the dominant contribution from Eq.(\ref{graph}) can be written as $f(\theta,\delta)=-140 \cos^2 \delta$.}
\end{figure}

\section{Calculation of the non-Gaussianity parameter $f_{NL}$}

The non-Gaussianity of a given theory of inflation and cosmological perturbations can be studied by looking at the expression of the non-linearity $f_{NL}$. 
This parameter is defined by the ratio of the bispectrum to the square of the isotropic part (as for example given in Eq.(\ref{piso})) of the power spectrum

\bea
\frac{6}{5}f_{NL}=\frac{B_{\zeta}(\vec{k}_{1},\vec{k}_{2},\vec{k}_{3})}{P_{\zeta}^{iso}(k_{1})P_{\zeta}^{iso}(k_{2})+perms.}.
\eea

As we saw in the previous sections, the complete expression for the bispectrum is made up of an `Abelian' and a `non-Abelian' parts, which can be read respectively from Eq.(\ref{class1}) and (\ref{equat1}). We are going to rewrite their sum below in a compact expression 

\bea\fl
B_{\zeta}(\vec{k}_{1},\vec{k}_{2},\vec{k}_{3})&=&B_{\zeta}^{A}(\vec{k}_{1},\vec{k}_{2},\vec{k}_{3})+B_{\zeta}^{NA}(\vec{k}_{1},\vec{k}_{2},\vec{k}_{3}).
\eea
Our next step will be to calculate the contributions from the two terms on the right hand side of the previous equation to the $f_{NL}$ parameter and establish a comparison between them.
The 'Abelian' part of the bispectrum was given in Eq.(\ref{class1}) as the contribution of three different pieces 

\bea
B_{\zeta}^{A}(\vec{k}_{1},\vec{k}_{2},\vec{k}_{3})=B_{\zeta}^{A_1}+B_{\zeta}^{A_2}+B_{\zeta}^{A_3},
\eea
where
\bea
B^{A_{1}}_{\zeta}\cong \left(N_{\phi}\right)^2 N_{\phi\phi}P^2,\\
B^{A_{2}}_{\zeta}\cong \left(N_{A}\right)^2 N_{AA}P^2 ,\\
B^{A_{3}}_{\zeta}\cong  N_{\phi} N_{A} N_{\phi A}P^2,
\eea
where the symbol $\cong$ stresses the fact that we are interested in the overall order of magnitude of the quantities above in terms of the parameters of the theory, without worrying about the details of numerical factors of order one. Also, we now indicate the power spectrum of the scalar field perturbation by $P$. At late times this is proportional to the power spectrum of the gauge fields. This can be seen by looking at the transverse mode function for $\delta B$ in Eq.(\ref{transverse-mode}) and at the longitudinal mode in Eq.(\ref{longitudinal-mode}), which, after taking the limit $x\rightarrow 0$, give $P_{+}=P$ and $P_{long}=2P$. We shortened $N^{i}_{a}$ by $N_{A}$, $N_{ab}^{ij}$ by $N_{AA}$ and $N_{\phi a}^{i}$ by $N_{\phi A}$ in order to make the expressions simpler, suppressing the gauge and the vector indices. Notice that in the theory we have chosen, no interactions occur between the scalar and the gauge fields, therefore we can set $N_{\phi A}\simeq 0$.\\
Similarly, we rewrite the 'non-Abelian' part of the bispectrum in a simplified fashion that leaves out numerical factors and complicated functions of the external momenta
\bea
B_{\zeta}^{NA}(\vec{k}_{1},\vec{k}_{2},\vec{k}_{3})\cong g_{c}^{2}H^{2} A \left(N_{A}\right)^3,
\eea
where we now call $A$ the background value of the spatial part of the gauge fields (suppressing all indices).\\
Let us now rewrite the isotropic part of the power spectrum
\bea
P_{\zeta}^{iso}\cong \left(N_{\phi}\right)^2 P\left(1+\beta\right),
\eea
where $\beta \equiv \left(N_{A}/N_{\phi}\right)^2$. \\
Before proceeding to the evaluation of $f_{NL}$, we need to make more assumptions about the scenario we are dealing with, since the derivatives of $N$ are model-dependent. 
Two possible choices are the vector curvaton model (introduced for the first time in \cite{Dimopoulos:2006ms}, see also  ~\cite{Dimopoulos:2008rf} for one of the possible realizations of this paradigm) and vector inflation~\cite{Golovnev:2008cf}. The former is the analogue of the (scalar) curvaton mechanism: a vector field plays the role of the curvaton, i.e. a field whose energy density is subdominant during inflation and, after the inflaton has decayed, it is is mainly responsible for the production of the curvature perturbation. In vector inflation, the universe exponential expansion is driven by vector fields, which need to be spatially oriented in such a way as to avoid large scale anisotropies; one possible field configuration is a triplet of orthogonal vector fields, another one consists in a large number of randomly oriented vector fields. In our model we are dealing with a triplet of vector fields, so if we assumed the fields were orthogonally oriented w.r.t one another, vector inflation would be a possible scenario to work in. A third possibility can be to consider a scalar field driven inflation in the presence of the $SU(2)$ gauge multiplet and to assume 
that the latter is undergoing slow-roll and its energy density is subdominant w.r.t. the energy density of the scalar inflaton. This way we keep the vector fields in the game, without imposing any restriction on their orientation in space and, at the same time, we 
avoid unwanted anisotropies in the power spectrum. We are going to consider both this last possibility, which we dare refer to as 'vector inflation' for simplicity, and the curvaton scenario. In 'vector inflation' $N_{a}^{i}$ is given by Eq.(\ref{derN}), from which we get 

\bea
N_{a}^{i}=\frac{A^{a}_{i}}{2m_{P}^{2}} ,\\
N_{ab}^{ij}=\frac{\delta_{ab}\delta^{ij}}{2m_{P}^{2}}. 
\eea
In the curvaton scenario~\cite{Lyth:2002my}

\bea
N_{a}^{i}=\frac{2}{3}r\frac{A_{i}^{a}}{\sum_{b}|\vec{A}^{b}|^2} ,\\\label{A_{tot}}
N_{ab}^{ij}=\frac{1}{3}r\frac{\delta_{ab}\delta^{ij}}{\sum_{c}|\vec{A}^{c}|^2} ,
\eea
where $r\equiv (3\rho_{A})/(3\rho_{A}+4\rho_{\phi})$, $\rho_{A}$ and $\rho_{\phi}$ being respectively the energy density of the vector fields and of the inflaton 
at the time where the vector field(s) decay.\\ 
We are now ready to evaluate $f_{NL}$ as the sum of three contributions corresponding to $B^{tree}_{\zeta,A_1}$, $B^{tree}_{\zeta,A_2}$ and $B^{tree}_{\zeta,NA}$

\bea
f_{NL}\cong f_{NL}^{A_1}+f_{NL}^{A_2}+f_{NL}^{NA},
\eea
where
\bea\label{fnl1}
f_{NL}^{A_1}&=&\frac{1}{(1+\beta)^2}\frac{N_{\phi\phi}}{N_{\phi}^2} ,\\\label{fnl2}
f_{NL}^{A_2}&=&\frac{\beta}{(1+\beta)^2}\frac{N_{AA}}{(N_{\phi})^2}  ,\\\label{fnl3}
f_{NL}^{NA}&=& \frac{\beta^2}{(1+\beta)^2}g_{c}^{2}\left(\frac{m}{H}\right)^2.
\eea
Numerical coefficients of order one were not reported in the previous equations. $m$ is a quantity with the dimensions of a mass and proportional to the Planck mass in the case of vector inflation and to $A_{tot}/\sqrt{r}$ (where we define $A_{tot}=\sqrt{\sum_{b}|\vec{A}^{b}|^2}$ from Eq.(\ref{A_{tot}})) in the vector curvaton model. \\
We summarize our results for the two models in Table 1, where we used the expressions $N_{\phi}\simeq (m_{P}\sqrt{\epsilon_{\phi}})^{-1}$ and $N_{\phi\phi}\simeq m_{P}^{-2}$, with $\epsilon_{\phi}\equiv(\dot{\phi}^2)/(2 m_{P}^{2}H^2)$. The background values $A$ and $A_{tot}$ together with $\epsilon_{\phi}$ and $H$ are all meant at time $x=x^{*}$.\\ 

\begin{table}[h]\centering
\caption{Order of magnitude of $f_{NL}$ in different scenarios.\\}
\begin{tabular}{|c||c|c|c|}\hline 
 $$ & $f_{NL}^{A_{1}}$ & $f_{NL}^{A_{2}}$ & $f_{NL}^{NA}$ \\
\hline
v.inflation &  $ \frac{\epsilon_{\phi}}{\left(1+\left(\frac{A}{m_{P}}\sqrt{\epsilon_{\phi}}\right)^2\right)^{2}}$ & $\frac{\epsilon_{\phi}^2}{\left(1+\left(\frac{A}{m_{P}}\sqrt{\epsilon_{\phi}}\right)^2\right)^{2}} \left(\frac{A}{m_{P}}\right)^2 $ & $\frac{\epsilon_{\phi}^2 g_{c}^{2}}{\left(1+\left(\frac{A}{m_{P}}\sqrt{\epsilon_{\phi}}\right)^2\right)^{2}} \left(\frac{A^2}{m_{P}H}\right)^2$ \\
\hline
v.curvaton & $\frac{\epsilon_{\phi}}{\left(1+\left(\frac{A m_{P}}{A_{tot}^2}\right)^{2}\epsilon_{\phi}r^2\right)^2}$ & $\frac{\epsilon_{\phi}^2 r^3}{\left(1+\left(\frac{A m_{P}}{A_{tot}^2}\right)^{2}\epsilon_{\phi}r^2\right)^2}\left(\frac{A m_{P}^{2}}{A_{tot}^{3}}\right)^2$ & $\frac{\epsilon_{\phi}^2 r^3 g_{c}^{2}}{\left(1+\left(\frac{A m_{P}}{A_{tot}^2}\right)^{2}\epsilon_{\phi} r^2\right)^2}\left(\frac{A^2 m_{P}^2}{A_{tot}^{3}H}\right)^2$ \\
\hline
\end{tabular}
\label{table1}
\end{table}

The ratios between the non-Abelian and the Abelian parts of the $f_{NL}$ parameter are
\bea\label{COM}
\frac{f_{NL}^{NA}}{f_{NL}^{A1}}\simeq\epsilon_{\phi}g_{c}^{2}\left(\frac{A}{H}\right)^2\left(\frac{A}{m_{P}}\right)^2  ,\,\,\,\,\,\,\,\,\,
\frac{f_{NL}^{NA}}{f_{NL}^{A2}}\simeq g_{c}^{2}\left(\frac{A}{H}\right)^2,
\eea
in vector inflation and 

\bea\label{COMP2}
\frac{f_{NL}^{NA}}{f_{NL}^{A1}}\simeq \frac{\beta^2 g_{c}^{2}}{r \epsilon_{\phi}} \left(\frac{A}{H}\right)^2 ,\,\,\,\,\,\,\,\,\,\,\,\,\,\,\,\,\,\,\,\,\,\,\,\,\,\,\,\,\,
\frac{f_{NL}^{NA}}{f_{NL}^{A2}} \simeq \frac{\beta g_{c}^{2}}{r^2 \epsilon_{\phi}} \left(\frac{A}{H}\right)^2 \left(\frac{A}{m_{P}}\right)^2 ,
\eea
for the vector curvaton.\\ 
When estimating the order of magnitude of $f_{NL}$ as calculated above, it is important to remember that our results in Eqs.~(\ref{fnl1}) through (\ref{COMP2}) 
and in Table 1 have been derived neglecting all vector and gauge indices. Also, these expressions involve the background value of the 
slowly-rolling fields $A_{i}^{a}$ in a multifield space and so $f_{NL}$ depends on the whole specific background configuration. In this respect, different possibilities can be considered and it turns out that, depending on the chosen configuration, the non-Abelian contributions to $f_{NL}$ can subdominant or dominant w.r.t. the Abelian ones. Notice that the ratios $g_{c}A/H=\left(g_{c}A/m_{0}\right)\left(m_{0}/H\right)$ for the different gauge fields are bounded from above, as shown in Eqs.~(\ref{cond1}) through (\ref{cond3}). There is a very large region of the parameter space of the background gauge field configurations for which this upper bound is much larger than one and, therefore, it does not prevent $f_{NL}^{NA}$ from dominating over $f_{NL}^{A_{1,2}}$. Here is an example: suppose that two of the gauge fields, for instance $\vec{B}^{1}$ and $\vec{B}^{2}$ are aligned, about equal in magnitude ($|\vec{B}^{1}|\simeq |\vec{B}^{2}|$) and that $|\vec{B}^{1,2}|\gg |\vec{B}^{3}|$. In this case, by looking at Eqs.~(\ref{cond1}) and (\ref{cond2}), it is easy to realize that the upper bounds can be much larger than one. Therefore, the ratios in Eqs.~(\ref{COM}) and (\ref{COMP2}) can be much bigger than one depending, for example, on how $A/H$ compares to the other quantities appearing in these expressions (notice: the theory puts no constraints on $A/H$ which can in principle be a very large number). Obviously,
inspecting Eqs.~(\ref{cond1}) through (\ref{cond3}), if the $A^a$'s are not approximately aligned and their magnitude $A$ is
roughly the same, then the upper bound $g_c\ll H/A$ holds.\\
As to the absolute value of $f_{NL}^{NA}$, expanding around $\beta \ll 1$, from Eq.(\ref{fnl3}) we have 

\bea\label{OOM}
f_{NL}^{NA}\simeq \beta^2 g_{c}^{2}\left(\frac{m_{P}}{H}\right)^2,\,\,\,\,\,\,\,\,\, f_{NL}^{NA}\simeq \frac{\beta^2 g_{c}^{2}}{r}\left(\frac{A}{H}\right)^2
\eea
respectively for vector inflation and the vector curvaton model. The ratio $m_{P}/H$ is of order $10^{9}$ and $A/H$ could be much bigger than one as well, so in principle $f_{NL}^{NA}$ can be much larger than one in both models. Equivalently, from Table 1, $f_{NL}^{NA}$ can be put in the following form (as a function of the slow-roll parameter $\epsilon_{\phi}$)

\[f_{NL}^{NA}\simeq \left\{ \begin{array}{ll}g_{c}^{2}\epsilon_{\phi}^{2}\left(\frac{A}{m_{P}}\right)^4\left(\frac{A}{H}\right)^2   &   \hspace{6ex} vector\,\,\,\, inflation \\g_{c}^{2}\epsilon^{2}_{\phi}r^3\left(\frac{m_{P}}{H}\right)^2\left(\frac{m_{P}}{A}\right)^2
&  \hspace{6ex} vector\,\,\,\, curvaton \end{array}\right.   \]

\section{Overview and conclusions}

We have considered a triplet of $SU(2)$ vector bosons non-minimally coupled to gravity, Eq.~(\ref{ac}), in order to have $M^2\simeq -2H^2$. Using the $\delta N$ and the Schwinger-Keldysh formalisms we have computed the contributions to the curvature perturbation three-point correlation function arising from the gauge fields self-interactions. These interactions are of two kinds, third-order interactions (proportional to one power of the $SU(2)$ coupling $g_{c}$) and fourth-order ones (proportional to $g_{c}^2$). The former provide a vanishing contribution to the bispectrum because of the antisymmetric properties of the Levi-Civita tensor appearing in the Lagrangian, the result being independent of the particular form of the wave functions appearing in the gauge bosons operator expansions. The quartic interactions produce instead a non-zero contribution, Eq.(\ref{equat1}). This result is anisotropic if the wavefunctions of the transverse and longitudinal modes of the vector fields are different from each other, which seems to be the case by looking at their equations of motion, Eqs.~(\ref{transverse}) and (\ref{longitudinal}).\\

The ongoing debate about the instability of the longitudinal mode sheds some doubts about the true physical meaning of the mode. Our analysis, in this respect, is pretty general, being mainly focused on studying the non-Gaussianity effects in non-Abelian theories. In addition to that, we tried to also use a more general approach when dealing with longitudinal modes, with a parametrization of the wavefunctions in terms of the transverse mode and of an unknown function $n(x)$, Eq.(\ref{approximation-n}), which, to a first approximation, can be set equal to a constant near horizon crossing, 
$n(x)=n(x^{*})$. A particular case of this parametrization occurs when the longitudinal and transverse modefunctions coincide, which offers an interesting result, i.e. the isotropization of the full bispectrum, both the `non-Abelian' and the `Abelian' parts, Eq.~(\ref{isotropic-bispectrum-res}) and~(\ref{class1}), respectively. We are aware that the particular Lagrangian we used represents only one possible model of primordial vector fields. Other theories have been proposed, which do not present any kind of instability (see for example, \cite{Dimopoulos:2007zb,Dimopoulos:2008yv,Yokoyama:2008xw,Dimopoulos:2009am}). \\

In the case of an anisotropic bispectrum we showed that it can be written in terms of isotropic parts, i.e. functions of the moduli of the external wave vectors $k_i$, 
modulated by anisotropic coefficients $I_{\alpha\beta\gamma}$ which depend on the angles between the (wave and gauge) vectors.
We studied the shape of the `isotropic parts' of the bispectrum, Eq.(\ref{equat1}) and Eq.(\ref{Kfirst}) through (\ref{Klast}), which turned out to peak in the local momenta configuration (see plots in Fig.1). Using this finding, we analysed the `anisotropic' part, Eq.~(\ref{anisotropic-coefficients1}) through (\ref{anisotropic-coefficients2}) for different spacial configurations of the gauge vectors. A limiting case is represented by the three components of the $SU(2)$ gauge group being all aligned with one another: because of the presence of the Levi-Civita tensors and with simple symmetry consideration, the contribution from (\ref{equat1}) to the bispectrum in this case is proven to be zero. This result could be interpreted as the Abelian limit of our theory, since the three components of the $SU(2)$ multiplet are all identifiable with a unique spatial direction.\\
Another example of vector fields configuration was provided in Eqs.(\ref{instance1})-(\ref{instance2}). For this particular case, a complete plot of the bispectrum was given in Fig.2 for the local configuration. It contains the information about the anisotropy of the bispectrum, showing how its amplitude is modulated according to the angles between the wavevectors $\vec{k}_i$ and the preferred directions induced by the vector fields.\\

Finally we have calculated the parameter $f_{NL}$, estimating the level of non-Gaussianity. We analyzed the different contributions to $f_{NL}$, i.e. the `Abelian', 
Eqs.~(\ref{fnl1}) and (\ref{fnl2}), and the `non-Abelian' ones, Eq.(\ref{fnl3}). We compared them both for vector inflation and for the vector curvaton models and for different background configurations. It turns out that one contribution can be dominant w.r.t. the other or viceversa, depending on the given background configuration, Eqs.~(\ref{COM}) and (\ref{COMP2}). Focusing on the order of magnitude of the `non-Abelian' contribution, we noticed that it can be much bigger than one, for a large region of parameter space, Eq.(\ref{OOM}).\\

In a forthcoming paper \cite{us}, we will study the trispectrum of the curvature 
perturbation in the same vector fields populated model, calculate its 
magnitude and shapes and and investigate the relationship between the $f_{NL}$ 
and $\tau_{NL}$ parameters of the theory.

\section*{Acknowledgments}
We would like to thank Massimo Pietroni for many useful discussions. This research has been partially supported by ASI contract I/016/07/0 "COFIS" and ASI contract Planck LFI Activity of Phase E2. 

\newpage

\vskip 1cm
\appendix
\setcounter{equation}{0}
\def\theequation{A.\arabic{equation}}
\vskip 0.2cm
\section{Background and first order perturbation equations for the gauge fields}\label{Background and first order perturbation equations for the gauge fields}

The equations of motion for the gauge fields have been completely derived for the $U(1)$ case in~\cite{Dimopoulos:2006ms}. We are going to carry out a similar calculation for the $SU(2)$ case  

\bea\label{eom}
\frac{1}{\sqrt{-g}}\p_{\mu}\Big[\sqrt{-g}g^{\mu\al}g^{\nu\b}\Big(F^{(AB)a}_{\al\b}+g_{c}\ep^{abc}B^{b}_{\al}B^{c}_{\b}\Big)\Big]+M^{2}g^{\mu\nu}B_{\mu}^{a}\nonumber\\+g_{c}\ep^{abc}g^{\ga\nu}g^{\de\b}F^{(AB)b}_{\ga\de}B^{c}_{\b}+g^{2}_{c}\ep^{abc}\ep^{bb'c'}g^{\nu\al}g^{\de\b}B^{c}_{\de}B^{b'}_{\al}B^{c'}_{\b}=0
\eea
where $F_{\mu\nu}^{(AB)a}\equiv\p_{\mu}B^{a}_{\nu}-\p_{\nu}B^{a}_{\mu}$.\\
The $\nu=0$ component of the equations of motion is

\bea\label{tc}
\p_{j}\dot{B}^{a}_{j}-\p_{j}\p_{j}B^{a}_{0}+a^{2}M^{2}B^{a}_{0}+g_{c}\ep^{abc}\Big[-\Big(\p_{j}B_{j}^{b}\Big)B^{c}_{0}-2B^{b}_{j}\p_{j}B_{0}^{c}-\dot{B}^{b}_{j}B^{c}_{j}\nonumber\\+g_{c}\ep^{cb'c'}B^{b}_{j}B^{b'}_{0}B^{c'}_{j}\Big]=0
\eea
where $B^{a}_{\mu}=B^{a}_{\mu}(\vec{x},t)$.\\
Let us now move to the spatial ($\nu=i$) part of (\ref{eom})
\bea\label{sc}
\ddot{B}_{i}^{a}+H\dot{B}_{i}^{a}-\frac{1}{a^{2}}\p_{j}\p_{j}B_{i}^{a}+M^{2}B_{i}^{a}-\p_{i}\dot{B}_{0}^{a}-H\p_{i}B_{0}^{a}+\frac{1}{a^{2}}\p_{i}\p_{j}B_{j}^{a}\nonumber\\
+g_{c}\ep^{abc}\Big[HB_{0}^{b}B_{i}^{c}+\dot{B}_{0}^{b}B_{i}^{c}+B_{0}^{b}\dot{B}_{i}^{c}\Big]-g_{c}\frac{\ep^{abc}}{a^2}\Big[\Big(\p_{j}B_{j}^{b})B^{c}_{i}+B_{j}^{b}\p_{j}B^{c}_{i}\Big]\nonumber\\
+g_{c}\ep^{abc}\Big[\Big(\p_{i}B_{0}^{b}\Big)B^{c}_{0}-\dot{B}^{b}_{i}B^{c}_{0}\Big]-g_{c}\frac{\ep^{abc}}{a^2}\Big[\Big(\p_{i}B_{j}^{b})B^{c}_{j}-\Big(\p_{j}B^{b}_{i}\Big)B_{j}^{c}\Big]\nonumber\\
+g_{c}^{2}\ep^{abc}\ep^{bb'c'}\Big[B^{c}_{0}B^{b'}_{i}B^{c'}_{0}\Big]-\frac{g_{c}^{2}}{a^2}\ep^{abc}\ep^{bb'c'}\Big[B^{c}_{j}B^{b'}_{i}B^{c'}_{j}\Big]=0
\eea
If we contract Eq.(\ref{eom}) with $\p_{\nu}$, we get the integrability condition 
\bea\label{e10}\fl
(aM)^2 \dot{B^{a}_{0}}-M^2\p_{i}B_{i}^{a}+3H\Big(\p_{i}\p_{i}B_{0}^{a}-\p_{i}\dot{B_{i}^{a}}\Big)+g_{c}\epsilon^{abc}\Big[2H\Big(\p_{i}B_{i}^{b}B_{0}^{c}+B_{i}^{b}\p_{i}B_{0}^{c}+\dot{B_{j}^{b}}B_{j}^{c}\Big)-\p_{j}B_{0}^{b}\dot{B}_{j}^{c}\nonumber\\\fl
-\ddot{B_{j}^{b}}B_{j}^{c}+\p_{j}\dot{B_{0}^{b}}B_{j}^{c}-\p^{2}B_{0}^{b}B_{0}^{c}-\p_{i}\dot{B_{i}^{b}}B_{0}^{c}+\frac{1}{a^2}\Big(B_{i}^{b}\p^{2}B_{i}^{c}+\p_{i}B_{j}^{b}\p_{j}B_{i}^{c}+B_{j}^{b}\p_{j}\p_{i}B_{i}^{c}+\p^{2}B_{j}^{b}B_{j}^{c}\nonumber\\\fl-\p_{i}\p_{j}B_{i}^{b}B_{j}^{c}\Big)\Big]+g_{c}^{2}\epsilon^{abc}\epsilon^{bb^{'}c^{'}}\Big[a^{2}\Big(\dot{B}_{0}^{c}B_{0}^{b^{'}}B_{0}^{c^{'}}+B_{0}^{c}\dot{B}_{0}^{b^{'}}B_{0}^{c^{'}}+B_{0}^{c}B_{0}^{b^{'}}\dot{B}_{0}^{c^{'}}\Big)+2HB_{i}^{c}B_{0}^{b^{'}}B_{i}^{c^{'}}\nonumber\\\fl-\dot{B}_{i}^{c}B_{0}^{b^{'}}B_{i}^{c^{'}}-B_{i}^{c}\dot{B}_{0}^{b^{'}}B_{i}^{c^{'}}-B_{i}^{c}B_{0}^{b^{'}}\dot{B}_{i}^{c^{'}}-\p_{i}B_{0}^{c}B_{i}^{b^{'}}B_{0}^{c^{'}}-B_{0}^{c}\p_{i}B_{i}^{b^{'}}B_{0}^{c^{'}}-B_{0}^{c}B_{i}^{b^{'}}\p_{i}B_{0}^{c^{'}}\nonumber\\\fl+\frac{1}{a^2}\Big(\p_{i}B_{j}^{c}B_{i}^{b^{'}}B_{j}^{c^{'}}+B_{j}^{c}\p_{i}B_{i}^{b^{'}}B_{j}^{c^{'}}+B_{j}^{c}B_{i}^{b^{'}}\p_{i}B_{j}^{c^{'}}\Big)\Big]=0
\eea
which reduces to Eq.($7$) of~\cite{Dimopoulos:2006ms} in the Abelian case.\\
Combining Eq.(\ref{e10}) with Eq.(\ref{tc}) we get

\bea
\fl
(aM)^2 \dot{B^{a}_{0}}-M^2\p_{i}B_{i}^{a}+3H\Big(a^{2}M^{2}B^{a}_{0}+g_{c}\ep^{abc}\Big[-\Big(\p_{j}B_{j}^{b}\Big)B^{c}_{0}-2B^{b}_{j}\p_{j}B_{0}^{c}-\dot{B}^{b}_{j}B^{c}_{j}+g_{c}\ep^{cb'c'}B^{b}_{j}B^{b'}_{0}B^{c'}_{j}\Big] \Big)\nonumber\\\fl+g_{c}\epsilon^{abc}\Big[2H\Big(\p_{i}B_{i}^{b}B_{0}^{c}+B_{i}^{b}\p_{i}B_{0}^{c}+\dot{B_{j}^{b}}B_{j}^{c}\Big)-\p_{j}B_{0}^{b}\dot{B}_{j}^{c}\nonumber\\\fl
-\ddot{B_{j}^{b}}B_{j}^{c}+\p_{j}\dot{B_{0}^{b}}B_{j}^{c}-\p^{2}B_{0}^{b}B_{0}^{c}-\p_{i}\dot{B_{i}^{b}}B_{0}^{c}+\frac{1}{a^2}\Big(B_{i}^{b}\p^{2}B_{i}^{c}+\p_{i}B_{j}^{b}\p_{j}B_{i}^{c}+B_{j}^{b}\p_{j}\p_{i}B_{i}^{c}+\p^{2}B_{j}^{b}B_{j}^{c}\nonumber\\\fl-\p_{i}\p_{j}B_{i}^{b}B_{j}^{c}\Big)\Big]+g_{c}^{2}\epsilon^{abc}\epsilon^{bb^{'}c^{'}}\Big[a^{2}\Big(\dot{B}_{0}^{c}B_{0}^{b^{'}}B_{0}^{c^{'}}+B_{0}^{c}\dot{B}_{0}^{b^{'}}B_{0}^{c^{'}}+B_{0}^{c}B_{0}^{b^{'}}\dot{B}_{0}^{c^{'}}\Big)+2HB_{i}^{c}B_{0}^{b^{'}}B_{i}^{c^{'}}\nonumber\\\fl-\dot{B_{i}^{c}}B_{0}^{b^{'}}B_{i}^{c^{'}}-B_{i}^{c}\dot{B_{0}^{b^{'}}}B_{i}^{c^{'}}-B_{i}^{c}B_{0}^{b^{'}}\dot{B}_{i}^{c^{'}}-\p_{i}B_{0}^{c}B_{i}^{b^{'}}B_{0}^{c^{'}}-B_{0}^{c}\p_{i}B_{i}^{b^{'}}B_{0}^{c^{'}}-B_{0}^{c}B_{i}^{b^{'}}\p_{i}B_{0}^{c^{'}}\nonumber\\\fl+\frac{1}{a^2}\Big(\p_{i}B_{j}^{c}B_{i}^{b^{'}}B_{j}^{c^{'}}+B_{j}^{c}\p_{i}B_{i}^{b^{'}}B_{j}^{c^{'}}+B_{j}^{c}B_{i}^{b^{'}}\p_{i}B_{j}^{c^{'}}\Big)\Big]=0
\eea
Plugging this into Eq.(\ref{sc}) we get

\bea\label{e9}\fl
\ddot{B}_{n}^{a}+H\dot{B}_{n}^{a}-\frac{1}{a^{2}}\p_{j}\p_{j}B_{n}^{a}+M^{2}B_{n}^{a}+2H\p_{n}B_{0}^{a}\nonumber\\\fl
-\frac{1}{\Big(aM\Big)^2}\p_{n}\Big[
-3H\Big(g_{c}\ep^{abc}\Big[-\Big(\p_{j}B_{j}^{b}\Big)B^{c}_{0}-2B^{b}_{j}\p_{j}B_{0}^{c}-\dot{B}^{b}_{j}B^{c}_{j}+g_{c}\ep^{cb'c'}B^{b}_{j}B^{b'}_{0}B^{c'}_{j}\Big] \Big)\nonumber\\\fl+g_{c}\epsilon^{abc}\Big[2H\Big(\p_{i}B_{i}^{b}B_{0}^{c}+B_{i}^{b}\p_{i}B_{0}^{c}+\dot{B_{j}^{b}}B_{j}^{c}\Big)-\p_{j}B_{0}^{b}\dot{B}_{j}^{c}\nonumber\\\fl
-\ddot{B_{j}^{b}}B_{j}^{c}+\p_{j}\dot{B_{0}^{b}}B_{j}^{c}-\p^{2}B_{0}^{b}B_{0}^{c}-\p_{i}\dot{B_{i}^{b}}B_{0}^{c}+\frac{1}{a^2}\Big(B_{i}^{b}\p^{2}B_{i}^{c}+\p_{i}B_{j}^{b}\p_{j}B_{i}^{c}+B_{j}^{b}\p_{j}\p_{i}B_{i}^{c}+\p^{2}B_{j}^{b}B_{j}^{c}\nonumber\\\fl-\p_{i}\p_{j}B_{i}^{b}B_{j}^{c}\Big)\Big]+g_{c}^{2}\epsilon^{abc}\epsilon^{bb^{'}c^{'}}\Big[a^{2}\Big(\dot{B_{0}^{c}}B_{0}^{b^{'}}B_{0}^{c^{'}}+B_{0}^{c}\dot{B}_{0}^{b^{'}}B_{0}^{c^{'}}+B_{0}^{c}B_{0}^{b^{'}}\dot{B}_{0}^{c^{'}}\Big)+2HB_{i}^{c}B_{0}^{b^{'}}B_{i}^{c^{'}}\nonumber\\\fl-\dot{B_{i}^{c}}B_{0}^{b^{'}}B_{i}^{c^{'}}-B_{i}^{c}\dot{B_{0}^{b^{'}}}B_{i}^{c^{'}}-B_{i}^{c}B_{0}^{b^{'}}\dot{B_{i}^{c^{'}}}-\p_{i}B_{0}^{c}B_{i}^{b^{'}}B_{0}^{c^{'}}-B_{0}^{c}\p_{i}B_{i}^{b^{'}}B_{0}^{c^{'}}-B_{0}^{c}B_{i}^{b^{'}}\p_{i}B_{0}^{c^{'}}\nonumber\\\fl+\frac{1}{a^2}\Big(\p_{i}B_{j}^{c}B_{i}^{b^{'}}B_{j}^{c^{'}}+B_{j}^{c}\p_{i}B_{i}^{b^{'}}B_{j}^{c^{'}}+B_{j}^{c}B_{i}^{b^{'}}\p_{i}B_{j}^{c^{'}}\Big)\Big]
\Big]\nonumber\\\fl
+g_{c}\ep^{abc}\Big[HB_{0}^{b}B_{n}^{c}+\dot{B}_{0}^{b}B_{n}^{c}+B_{0}^{b}\dot{B}_{n}^{c}\Big]-g_{c}\frac{\ep^{abc}}{a^2}\Big[\Big(\p_{j}B_{j}^{b})B^{c}_{n}+B_{j}^{b}\p_{j}B^{c}_{n}\Big]\nonumber\\\fl
+g_{c}\ep^{abc}\Big[\Big(\p_{n}B_{0}^{b}\Big)B^{c}_{0}-\dot{B}^{b}_{n}B^{c}_{0}\Big]-g_{c}\frac{\ep^{abc}}{a^2}\Big[\Big(\p_{n}B_{j}^{b})B^{c}_{j}-\Big(\p_{j}B^{b}_{n}\Big)B_{j}^{c}\Big]\nonumber\\\fl
+g_{c}^{2}\ep^{abc}\ep^{bb'c'}\Big[B^{c}_{0}B^{b'}_{n}B^{c'}_{0}\Big]-\frac{g_{c}^{2}}{a^2}\ep^{abc}\ep^{bb'c'}\Big[B^{c}_{j}B^{b'}_{n}B^{c'}_{j}\Big]=0.
\eea
Let us consider the background part of the vector fields, i.e. $\p_{i}B_{\mu}^{a}=0$. Then from Eq.(\ref{tc})
\bea\label{backzero}
a^{2}M^{2}B^{a}_{0}+g_{c}\ep^{abc}\Big[-\dot{B}^{b}_{j}B^{c}_{j}+g_{c}\ep^{cb'c'}B^{b}_{j}B^{b'}_{0}B^{c'}_{j}\Big]=0.
\eea
Before proceeding with the derivation of the equations of motion for the background and the field perturbations, it is necessary to make some comments on Eqs.~(A.6) and (A.7). One approximation that we have been using in this paper is allowing the $A_{i}^{a}$ fields to undergo slow-roll during inflation. One possible way to achieve this is by restricting the parameter space of the background gauge fields through the request that their temporal components should be much smaller than the spatial ones, $B^{b}_{0}\ll |B^{b}_{i}|/a(t)$, $b=1,2,3$, and, in addition to that, assuming $B_{0}^{b}\simeq B_{0}^{c}$, $b,c=1,2,3$. With these assumptions, the temporal component can be factored out in Eq.~(\ref{backzero}), using the approximation $\dot{B}^{a}_{i}\simeq H B^{a}_{i}$ (valid in a slow-roll regime). A solution to (\ref{backzero}) is then given by $B_{0}=0$. Adopting this solution and plugging it in Eq~(A.6), it is easy to show that a slow-roll equation of motion for the physical fields
\bea
\ddot{A}_{i}^{a}+3H\dot{A}_{i}^{a}+m_{0}^{2}A_{i}^{a}=0
\eea  
follows from (A.6) if $\dot{H}\ll m_{0}^{2}$ and 

\bea\label{cond1}\fl
\left(\frac{g_{c}A^{1}}{m_{0}}\right)^2\ll\left|\frac{\left(A^{1}\right)^2}{\left(A^{2}\right)^2+\left(A^{3}\right)^2-\left(A^{3}\right)^2\cos^2\theta_{13}-\left(A^{2}\right)^2\cos^2\theta_{12}}\right|,\\\label{cond2}\fl
\left(\frac{g_{c}A^{2}}{m_{0}}\right)^2\ll\left|\frac{\left(A^{2}\right)^2}{\left(A^{1}\right)^2+\left(A^{3}\right)^2-\left(A^{3}\right)^2\cos^2\theta_{23}-\left(A^{1}\right)^2\cos^2\theta_{12}}\right|,\\\label{cond3}\fl
\left(\frac{g_{c}A^{3}}{m_{0}}\right)^2\ll\left|\frac{\left(A^{3}\right)^2}{\left(A^{1}\right)^2+\left(A^{2}\right)^2-\left(A^{2}\right)^2\cos^2\theta_{23}-\left(A^{1}\right)^2\cos^2\theta_{13}}\right|,
\eea
are satisfied. In the equations above, we defined $A^{a}\equiv|\vec{A}^{a}|$ and $\cos\theta_{ab}\equiv\hat{A}^{a}\cdot\hat{A}^{b}$, $a$ and $b$ running over the gauge indices. The quantities appearing on the right-hand sides of Eqs.(\ref{cond1}) through (\ref{cond3}) can be either large or small w.r.t. one, depending on the specific background configuration, i.e. on the moduli of the gauge fields and the angles $\theta_{ab}$.\\
Suppose now the conditions described above are all met, then from Eq~(A.6), in terms of the comoving fields, we have
\bea\label{eomB}
\ddot{B_{i}^{a}}+H\dot{B^{a}_{i}}+M^2B_{i}^{a}=0. 
\eea
Let us now derive the equations for the perturbations. Eq.(\ref{tc}) becomes

\bea\label{sc13}\fl
\p_{j}\de \dot{B}_{j}^{a}-\p^2\de B_{0}^{a}+a^2 M^2\de B_{0}^{a}+g_{c}\ep^{abc}\Big[-\p_{j}\de B_{j}^{b}B_{0}^{c}-2B_{j}^{b}\p_{j}\de B_{0}^{c}-\de \dot{B}_{j}^{b}B_{j}^{c}-\dot{B}_{j}^{b}\de B_{j}^{c}\nonumber\\\fl+g_{c}\ep^{cb^{'}c^{'}}\Big(\de B_{j}^{b} B_{0}^{b^{'}} B_{j}^{c^{'}}+ B_{j}^{b} \de B_{0}^{b^{'}} B_{j}^{c^{'}}+ B_{j}^{b} B_{0}^{b^{'}} \de B_{j}^{c^{'}}\Big)\Big] =0
\eea
Eq.(\ref{eom}) for the field perturbations gives

\bea\label{sc1}\fl
\de \ddot{B}_{i}^{a}+H\de\dot{B}_{i}^{a}-\frac{1}{a^{2}}\p_{j}\p_{j}\de B_{i}^{a}+M^{2}\de B_{i}^{a}+\frac{1}{a^{2}}\p_{i}\p_{j}\de B_{j}^{a}-H\p_{i}\de B_{0}^{a}-H\p_{i}\de B_{0}^{a}\nonumber\\\fl
-\frac{g_{c}}{a^2}\ep^{abc}\Big[\Big(\p_{j}\de B_{j}^{b})B^{c}_{i}+B_{j}^{b}\p_{j}\de B^{c}_{i}\Big]
-\frac{g_{c}}{a^2}\ep^{abc}\Big[\Big(\p_{i}\de B_{j}^{b})B^{c}_{j}-\Big(\p_{j}\de B^{b}_{i}\Big)B_{j}^{c}\Big]\nonumber\\\fl
-\frac{g_{c}^{2}}{a^2}\ep^{abc}\ep^{bb'c'}\Big[\de B^{c}_{j}B^{b'}_{i}B^{c'}_{j}+B^{c}_{j}\de B^{b'}_{i}B^{c'}_{j}+B^{c}_{j}B^{b'}_{i}\de B^{c'}_{j}\Big]\nonumber\\\fl
+g_{c}\ep^{abc}\Big[H\left(B^{b}_{0}\DB^{c}_{i}+\DB_{0}^{b}B_{i}^{c}\right)+\dot{\DB^{b}_{0}}B_{i}^{c}+\dot{B^{b}_{0}}\DB_{i}^{c}+\DB^{b}_{0}\dot{B_{i}^{c}}+B^{b}_{0}\dot{\DB^{c}_{i}}\Big]\nonumber\\\fl
+g_{c}\ep^{abc}\Big[\p_{i}\DB^{b}_{0}\DB_{0}^{c}-\dot{\DB_{i}^{b}}B_{0}^{c}-\dot{B_{i}^{b}}\DB^{c}_{0}\Big]\nonumber\\\fl
+g_{c}^{2}\ep^{abc}\ep^{bb^{'}c^{'}}\Big[\DB_{0}^{c}B_{i}^{b^{'}}B_{0}^{c^{'}}+B_{0}^{c}\DB_{i}^{b^{'}}B_{0}^{c^{'}}+B_{0}^{c}B_{i}^{b^{'}}\DB_{0}^{c^{'}}\Big]=0
\eea
Finally from Eq.(\ref{e9}) we get

\bea\label{perturbationsEQ}\fl
\ddot{\DB^{a}_{i}}+H\dot{\DB^{a}_{i}}-\frac{1}{a^2}\p^{2}\DB^{a}_{i}+M^2\DB^{a}_{i}+2H\p_{i}\DB^{a}_{0}+(\sim g_{c}terms)=0.
\eea
When calculating $n-point$ functions for the gauge bosons, the eigenfunctions we need are provided by free-field solutions, i.e. by solutions of Eq.(\ref{perturbationsEQ}) with $g_{c}$ being set to zero. This is exactly the Abelian limit, in fact in this case Eq.(\ref{perturbationsEQ}) corresponds to (18) of~\cite{Dimopoulos:2006ms} and can be decomposed into a transverse and a longitudinal part

\bea\label{transverse}
\left[\p_{0}^{2}+H\p_{0}+M^2+\left(\frac{k}{a}\right)^2\right]\delta\vec{B}^{T}=0 \\\label{longitudinal}
\left[\p_{0}^{2}+\left(1+\frac{2k^2}{k^2+\left(aM\right)^2}\right)H\p_{0}+M^2+\left(\frac{k}{a}\right)^2\right]\delta\vec{B}^{||}=0
\eea
where the time derivatives are intended w.r.t. cosmic time.

\def\theequation{B.\arabic{equation}}
\vskip 0.2cm

\section{Calculation of the number of e-foldings of single-(scalar)field driven inflation in the presence of a vector multiplet}
\label{Calculation of the number of e-foldings of single-(scalar)field driven inflation in the presence of a vector multiplet}

Let us consider the complete Lagrangian of our theory as in Eq.(\ref{ac}). Let us assume that the $SU(2)$ gauge multiplet undergoes slow-roll as well as the scalar field but the latter provides the dominant part of the energy density of the universe. This last hypothesis is necessary in order to produce isotropic inflation (i.e. in order for the anisotropy in the expansion that the vector fields introduce to be negligible w.r.t. the isotropic contribution from the scalar field). The expression of the number of e-foldings is 

\bea\label{N__}
N=N_{scalar}+N_{vector}=N_{scalar}+\frac{1}{4m_{P}^{2}}\sum_{a=1,2,3}\vec{A}^{a}\cdot\vec{A}^{a}.
\eea
The previous expression can be easily derived from the equations of motion of the system neglecting terms that are proportional to the $SU(2)$ coupling constant $g_{c}$ and assuming slow-roll conditions for both the scalar the gauge fields.\\
The starting point is represented by Einstein equations

\bea\label{EEdeltaN}
H^2=\frac{8 \pi G}{3}\left(\rho_{scalar}+\rho_{vector}\right).
\eea
where we split the energy density into a scalar and a vector contribution. In slow-roll approximation, $\rho_{scalar}\sim V(\phi)$. Let us  calculate $\rho_{vector}$. The energy momentum tensor for the gauge bosons

\bea
T_{\mu\nu}^{vector}=2\frac{\delta \textit{L}}{\delta g^{\mu\nu}}-g_{\mu\nu}\textit{L}
\eea
where, as a remainder, $\textit{L}=-(1/4)g^{\mu\alpha}g^{\nu\beta}F_{\mu\nu}^{a}F^{a}_{\alpha\beta}+(M^2/2)g^{\mu\nu}B^{a}_{\mu}B^{a}_{\nu}$. So we get

\bea
T_{00}^{vector}&=&\frac{\dot{B}^{a}_{i}\dot{B}^{a}_{i}}{2 a^2}+\frac{m_{0}^2}{2a^2} B^{a}_{i} B^{a}_{i}+\frac{m_{0}^2}{2} B^{a}_{0} B^{a}_{0}-\frac{H}{a^2}\dot{B}^{a}_{i}{B}^{a}_{i}+\frac{H^2}{2a^2}{B}^{a}_{i}{B}^{a}_{i}\nonumber\\&+&\frac{g_{c}}{a^2}\ep^{abc}\dot{B}^{a}_{i}B_{0}^{b}B_{i}^{c}+\frac{g_{c}^{2}}{2a^2}\ep^{abc}\ep^{ab^{'}c^{'}}B_{0}^{b}B_{i}^{c}B_{0}^{b^{'}}B_{i}^{c^{'}}\nonumber\\&+&\frac{g_{c}^{2}}{4a^4}\ep^{abc}\ep^{ab^{'}c^{'}}B^{b}_{i}B^{c}_{j}B^{b^{'}}_{i}B^{c^{'}}_{j}
\eea
where sums are taken over all repeated indices. Let us write this in terms of the physical fields 

\bea\fl
T_{00}^{vector}&=&\frac{\dot{A}_{i}^{a}\dot{A}_{i}^{a}}{2}+\frac{m_{0}^2}{2}\left({A}_{i}^{a}{A}_{i}^{a}+A_{0}^{a}A_{0}^{a}\right)+g_{c}\ep^{abc}\left(HA_{i}^{a}+\dot{A}^{a}_{i}\right)A^{b}_{0}A^{c}_{i}+\frac{g_{c}^{2}}{2}\ep^{abc}\ep^{ab^{'}c^{'}}A_{0}^{b}A_{i}^{c}A_{0}^{b^{'}}A_{i}^{c^{'}}\nonumber\\\fl&+&\frac{g_{c}^2}{4}\ep^{abc}\ep^{ab^{'}c^{'}}{A}_{i}^{b}{A}_{j}^{c}{A}_{i}^{b^{'}}{A}_{j}^{c^{'}}
\eea
If we neglect the non-Abelian contribution and we set $A^{a}_{0}=0$, we are left with the Abelian result~\cite{Golovnev:2008cf}

\bea\label{naro}\fl
T_{00}^{vector}&=&\frac{\dot{A}_{i}^{a}\dot{A}_{i}^{a}}{2}+\frac{m_{0}^2}{2}{A}_{i}^{a}{A}_{i}^{a}
\eea

The equation of motion for the background vector multiplet $\vec{A}^{a}$ can be derived from Eq.(\ref{eomB})

\be\label{eom-de-sitter}
\ddot{A_{i}^{a}}+3H\dot{A^{a}_{i}}+m_{0}^2A_{i}^{a}=0. 
\ee
which is equal to the equation of a light scalar field of mass $m_{0}$, if $m_{0} \ll H$. If the conditions for accelerated expansions are met, Eq.(\ref{eom-de-sitter}) reduces to 

\bea
3H\dot{A^{a}_{i}}+m_{0}^2A_{i}^{a}\sim 0. 
\eea
We are now ready to derive Eq.(\ref{N__}). Let us start from the definition of N and keep in mind Eq.(\ref{EEdeltaN}), where we are assuming the existence of a scalar fields $\phi$ in de-Sitter with a separable potential governed by the usual (background) equation 

\bea
\ddot{\phi}+3H\dot{\phi}+V^{'}=0
\eea
and slowly rolling down their potential. Then we have

\bea
N&=&\int_{t^{*}}^{t} Hdt^{'}=\int_{t^{*}}^{t} H^2\frac{dt^{'}}{H}=8 \pi G\int_{t^{*}}^{t}\frac{V(\phi)}{3H}dt^{'}+8 \pi G\int_{t^{*}}^{t}\frac{V(A)}{3H}dt^{'}\nonumber\\&=&8 \pi G\int_{t^{*}}^{t}\frac{V(\phi)}{3H}\frac{dt}{d\phi}d\phi+8 \pi G\sum_{a}\int_{t^{*}}^{t}\left(\frac{m_{0}^2}{2}\right)\frac{{A}_{i}^{a}{A}_{i}^{a}}{3H}\frac{dA^{a}dt^{'}}{dA^{a}}
\eea
where $A^{a}\equiv \vec{A^{a}}\cdot\vec{A^{a}}$. So

\bea\label{N___}
N&=&8 \pi G\int_{\phi(t^{*})}^{\phi(t)}\frac{V(\phi)}{3H\dot{\phi}}d\phi+8 \pi G\sum_{a}\int_{A^{a}(t^{*})}^{A^{a}(t)}\left(\frac{m_{0}^2}{4}\right)\frac{{A}_{i}^{a}{A}_{i}^{a}}{3H\dot{A}^{a}_{j}A^{a}_{j}}dA^{a}\nonumber\\&=&-\frac{1}{m_{P}^{2}}\int_{\phi(t^{*})}^{\phi(t)}\frac{V}{V^{'}}d\phi+\frac{1}{m_{P}^{2}}\sum_{a}\int_{A^{a}(t^{*})}^{A^{a}(t)}\left(\frac{m_{0}^2}{4}\right)\frac{{A}_{i}^{a}{A}_{i}^{a}}{(-m_{0}^2){A}^{a}_{j}A^{a}_{j}}dA^{a}\nonumber\\&=&-\frac{1}{m_{P}^{2}}\int_{\phi(t^{*})}^{\phi(t)}\frac{V}{V^{'}}d\phi-\left(\frac{1}{4 m_{P}^{2}}\right)\sum_{a}\int_{A^{a}(t^{*})}^{A^{a}(t)} dA^{a}
\eea
after using the slow-roll conditions. Eq.(\ref{N__}) is thus recovered. \\
In the final expression for the bispectrum then we can substitute

\bea\label{derN}
N^{i}_{a}\equiv\frac{dN}{dA^{a}_{i}}=\left(\frac{1}{2m_{P}^{2}}\right)A^{a}_{i}
\eea
where the derivatives are as usual calculated at the initial time $\eta^{*}$.\\
The upper limits in integrals such as the ones in Eq.(\ref{N___}) depend on the chosen path in field space and so they also depend on the initial field configuration. It is important to notice though, as also stated in~\cite{Vernizzi:2006ve}, that if the final time is chosen to be approaching (or later than) the end of inflation, the fields are supposed to have reached their equilibrium values and so $N$ becomes independent of the field values at the final time $t$. Eq.(\ref{derN}) is thus recovered.

\def\theequation{C.\arabic{equation}}
\vskip 0.2cm
\section{Complete expressions for the functions appearing in the bispectrum from quartic interactions}
\label{Complete expressions for the functions appearing in the bispectrum from quartic interactions}

We list below the complete expressions for the functions $A_{\alpha\beta\gamma}$, $B_{\alpha\beta\gamma}$ and $C_{\alpha\beta\gamma}$ ($\alpha,\beta,\gamma=R,l$) appearing in Eq.(\ref{Kfirst}) through (\ref{Klast})

\bea\fl
A_{EEE}&\equiv& kx^{*2}\Big(-k^2(k^3_1+k^3_2+k^3_3-4k_1k_2k_3)-k^3(k_2k_3+k_1 k_2+k_1 k_3)\nonumber\\\fl&&+k_1k_2k_3(k^2_1+k^2_2+k^2_3-k_2k_3-k_1 k_2-k_1 k_3)x^{*2}\Big) \\
\fl
B_{EEE}&\equiv& \Big(k^3_1+k^3_2+k^3_3\Big)x^{*3}\Big(-k^3+k_1k_2k_3 x^{*2}\Big)\\
\fl
C_{EEE}&\equiv& -k\Big(k^3_1+k^3_2+k^3_3\Big)x^{*2}\Big(-k^2+(k_2k_3+k_1 k_2+k_1 k_3)x^{*2}\Big)\\
\fl
A_{lll}&\equiv& 16k^6\Big[2k^3(k_1+k_2)(k_1+k_3)(k_2+k_3)-3\Big(k_1^4(k_2+k_3)^2+k_2^2k_3^2(k_2+k_3)^2\nonumber\\\fl&&+k_1^3(2k_2^3+9k_2^2k_3+9k_2k_3^2+2k_3^3)+k_1k_2k_3(2k_2^3+9k_2^2k_3+9k_2k_3^2+2k_3^3)\nonumber\\\fl&&+k_1^2(k_2^4+9k_2^3k_3+19k_2^2k_3^2+9k_2k_3^3+k_3^4)\Big)\Big]x^{*2}-8k^5\Big[-3kk_1^2k_2^2k_3^2\nonumber\\\fl&&+2k^4(k_1+k_2)(k_1+k_3)(k_2+k_3)+2k^2k_1k_2k_3(k_2k_3+k_1k_2+k_1k_3)\nonumber\\\fl&&-2k^3(k_2k_3+k_1k_2+k_1k_3)^2+6k_1k_2k_3(k_1^3(k_2+k_3)+k_2k_3(k_2+k_3)^2\nonumber\\\fl&&+2k_1^2(k_2^2+3k_2k_3+k_3^2)+k_1(k_2^3+6k_2^2k_3+6k_2k_3^2+k_3^3))-3k\Big(k_1^4(k_2+k_3)^2\nonumber\\\fl&&+k_2k_3(k_2+k_3)^2+k_1^3\Big(2k_2^3+9k_2^2k_3+9k_2k_3^2+2k_3^3\Big)+k_1k_2k_3\nonumber\\\fl&&\times\Big(2k_2^3+9k_2^2k_3+9k_2k_3^2+2k_3^3\Big)+k_1^2(k_2^4+9k_2^3k_3+19k_2^2k_3^2+9k_2k_3^3+k_3^4)\Big)\Big]x^{*4}\nonumber\\\fl&&+4k^2\Big[-3k_2^4k_3^4(k_2+k_3)^2-3k_1^6(k_2+k_3)^4-3k_1^5(k_2+k_3)^2\nonumber\\\fl&&\times(2k_2^3+11k_2^2k_3+2k_3^3+k_2k_3(-2k+11k_3))\nonumber\\\fl&&-3k_1^4(k_2^6+15k_2^5k_3+k_3^6+k_2k_3^4(-4k+15k_3)+2k_2^3k_3^2(-8k+47k_3)\nonumber\\\fl&&+k_2^2k_3(-4k+61k_3)+k_2^2k_3^2(k^2-16kk_3+61k_3^2))-3k_1^3k_2k_3(k_2+k_3)\nonumber\\\fl&&\times\Big(4k_2^4+2k^2k_2k_3+31k_2^3k_3+63k-2^2k_3^2+31k_2k_3^3+4k_3^4\nonumber\\\fl&&-2k(k_2^3+7k_2^2k_3+7k_2k_3^2+k_3^3)\Big)-k_1k_2^2k_3^2\Big(-2k^5-6kk_2k_3(k_2+k_3)^2\nonumber\\\fl&&+3k_2k_3(4k_2^3+15k_2^2k_3+15k_2k_3^2+4k_3^3)\Big)-k_1^2k_2k_3\Big(2k^4k_2k_3-2k^5(k_2+k_3)\nonumber\\\fl&&+3k^2k_2k_3(k_2+k_3)^2-12kk_2k_3(k_2^3+4k_2^2k_3+4k_2k_3^2+k_3^3)\nonumber\\\fl&&+3k_2k_3(6k_2^4+35k_2^3k_3+61k_2^2k_3^2+35k_2k_3^3+6k_3^4)\Big)\Big]x^{*6}+6k_1^2k_2^2k_3^2\Big[k_1^4(k_2+k_3)^2\nonumber\\\fl&&+k_2^2k_3^2(k_2+k_3)^2+k_1^3(2k_2^3+9k_2^2k_3+9k_2k_3^2+2k_3^3)+k_1k_2k_3\nonumber\\\fl&&\times\Big(2k_2^3+9k_2^2k_3+9k_2k_3^2+2k_3^3\Big)+k_1^2(k_2^4+9k_2^3k_3+19k_2^2k_3^2+9k_2k_3^3+k_3^4)\nonumber\\\fl&&+k^2(k_2k_3+k_1k_2+k_1k_3)^2-2k\Big(k_2^2k_3^2(k_2+k_3)+k_1^3(k_2+k_3)^2\nonumber\\\fl&&+2k_1k_2k_3(k_2^2+3k_2k_3+k_3^2)+k_1^2(k_2^3+6k_2^2k_3+6k_2k_3^2+k_3^3)\Big)\Big]x^{*8}\nonumber\\\fl&&-3k_1^4k_2^4k_3^4x^{*10}\\
\fl
B_{lll}&\equiv& 8k^4\Big(k_1^3+k_2^3+k_3^3\Big)x^{*3}\Big(4k^4(k_1+k_2+k_3)-2k^2(k_1+k_2)(k_1+k_3)\nonumber\\\fl&&\times(k_2+k_3)x^{*2}+k_1k_2k_3(k_2k_3+k_1k_2+k_1k_3)x^{*4}\Big)\\
\fl
C_{lll}&\equiv&-4k^3\Big(k^3_1+k^3_2+k^3_3\Big)x^{*2}\Big(8k^6-4k^4(k_1+k_2+k_3)^2x^{*2}+2k^2\Big(k_2k_3+k_1k_2\nonumber\\\fl&&+k_1k_3\Big)^2x^{*4}-k_1^2k_2^2k_3^2x^{*6}\Big) \\
\fl
A_{llE}&\equiv& 4k^5\Big[2k^3(k_1+k_2)(k_1+k_2+2k_3)-2k^2\Big(k_1^3+k_2^3+4k_2^2k_3\nonumber\\\fl&&-2k_3^3+4k_1^2(k_2+k_3)+4k_1k_2(k_2+2k_3)\Big)+3k_1k_2\Big(2k_2k_3(k_2+k_3)\nonumber\\\fl&&+k_1^2(k_2+2k_3)+k_1(k_2^2+6k_2k_3+2k_3^2)\Big)\Big]x^{*2}\nonumber\\\fl&&-2k^3\Big[2k^3k_1k_2(2k_2k_3+k_1k_2+2k_1k_3)-2k^2\Big(k_1^4(k_2+k_3)\nonumber\\\fl&&+k_2^2k_3(k_2^2+2k_2k_3-2k_3^2)+k_1^3(3k_2^2+6k_2k_3+2k_3^2)\nonumber\\\fl&&+k_1k_2(k_2^3+6k_2^2k_3+4k_2k_3^2-4k_3^3)+k_1^2(3k_2^3+14k_2^2k_3+4k_2k_3^2-2k_3^3)\Big)\nonumber\\\fl&&+3k_1k_2(k_1+k_2)(k_1^3(k_2+2k_3)+2k_2k_3(k_2^2+3k_2k_3+2k_3^2)\nonumber\\\fl&&+2k_1^2(k_2^2+5k_2k_3+3k_3^2)+k_1(k_2^3+10k_2^2k_3+16k_2k_3^2+4k_3^3))\Big]x^{*4}\nonumber\\\fl&&+kk_1^2k_2^2\Big[3k_1^3(k_2+2k_3)^2+2k_3\Big(-3kk_2^2k_3+6k_2^2k_3(k_2+k_3)\nonumber\\\fl&&-k^2(k_2^2+2k_2k_3-2k_3^2)\Big)-2k_1k_3\Big(2k^2(k_2+k_3)+3kk_2(k_2+2k_3)\nonumber\\\fl&&-3k_2(2k_2^2+9k_2k_3+4k_3^2)\Big)+k_1^2\Big(3k_2^3+30k_2^2k_3-6k_2(k-9k_3)k_3\nonumber\\\fl&&+2k_3(-k^2-3kk_3+6k_3^2)\Big)\Big]x^{*6}+3k_1^4k_2^4k_3^2 x^{*8} \\
\fl
B_{llE}&\equiv& -2k^2(k_1^3+k_2^3-2k_3^3)\Big(4k^4k_1+4k^4k_2+4k^4k_3+\Big[-2k^2(k_1+k_2)\nonumber\\\fl&&\times(k_2k_3+k_1k_2+k_1k_3)\Big]x^{*5}\nonumber\\\fl&&+k_1^2k_2^2k_3 x^{*7}\Big)\\
\fl
C_{llE}&\equiv& 2k^3(k_1^3+k_2^3-2k_3^3)\Big[4k^4 x^{*2}-2k^2(k_1+k_2)(k_1+k_2+2k_3)x^{*4}\nonumber\\\fl&&+k_1k_2\Big(2k_2k_3+k_1(k_2+2k_3)\Big)x^{*6}\Big]\\
\fl
A_{EEl}&\equiv& 2k^4\Big[-2k_1^4-2k_2^4-2k_2^3k_3-k^2k_3^2+3k_2^2k_3^2+k_3^4-2k_1^3(k_2+k_3)\nonumber\\\fl&&+3k_1^2k_3(2k_2+k_3)+k_2k_3(-2k^2+kk_3+4k_3^2)\nonumber\\\fl&&+k_1\Big(-2k_2^3+6k_2^2k_3+9k_2k_3^2+4k_3^3-2k^2(k_2+k_3)\nonumber\\\fl&&+kk_3(2k_2+k_3)\Big)\Big]+k^2k_3\Big[2k_1^3k_2k_3+2k_1^4(2k_2+k_3)\nonumber\\\fl&&-3k_1^2k_3(2k_2^2+2k_2k_3+k_3^2)+k_2k_3(2k_2^3-3k_2k_3^2-k_3^3)\nonumber\\\fl&&+k_1(4k_2^4+2k^2k_2k_3+2k_2^3k_3-6k_2^2k_3^2-5k_2k_3^3-k_3^4)\Big]x^{*2}\nonumber\\\fl&&-3k_1^2k_2^2k_3^4 x^{*4}  \\
\fl
B_{EEl}&\equiv&-k^2(2k_1^3+2k_2^3-k_3^3)\Big(2k^3 x^{*}+\Big[-2k_1k_2k_3-k_1k_3^2-k_2k_3^2\Big]x^{*3}\Big) \\
\fl
C_{EEl}&\equiv& k(2k_1^3+2k_2^3-k_3^3)\Big(2k^4+\Big[-2k^2k_1k_2-2k^2k_1k_3-2k^2k_2k_3-k^2k_3^2\Big]x^{*2}\nonumber\\\fl&&+k_1k_2k_3^2x^{*4}\Big)
\eea
where $k\equiv k_{1}+k_{2}+k_{3}$.

\def\theequation{D.\arabic{equation}}
\vskip 0.2cm
\section{Profile of the bispectrum: detailed expressions of the plotted functions}
\label{Profile of the bispectrum: detailed expressions of the plotted functions}

We are going to study the profile of the 'non-Abelian' contribution to the bispectrum in terms of the angles between gauge and wave vectors in the local configuration and for the set up considered in Eq.(\ref{instance1}) and (\ref{instance2}). The coefficients $I_{\alpha\beta\gamma}$ then become

\bea\fl
I_{EEE}&=&m^2 N^{4}_{A}\Big[-20-24\cos\delta+2\cos\theta-12\cos^2\delta+12\cos^2\theta-2\cos^3\theta+6\cos\theta\cos^2\delta\nonumber\\\fl&&-2\cos^2\theta\cos\delta+2\cos\delta\cos^3\theta\Big] ,\\\fl  
I_{lll}&=& m^2 N^{4}_{A}\Big[4\cos^2\delta\Big],\\\fl
I_{llE}&=& m^2 N^{4}_{A}\Big[4-2\cos\theta-6\cos^2\theta-4\cos^2\delta\Big],\\\fl
I_{lEl}&=&m^2 N^{4}_{A}\Big[-4\cos^2\delta \Big],\\\fl
I_{Ell}&=&m^2 N^{4}_{A}\Big[-4\cos^2\delta\Big] ,\\\fl
I_{EEl}&=&m^2 N^{4}_{A}\Big[ -2\cos^2\delta\Big],\\\fl
I_{ElE}&=& m^2 N^{4}_{A}\Big[4-4\cos^2\theta-8\cos^2\delta\Big],\\\fl
I_{lEE}&=& m^2 N^{4}_{A}\Big[2+\cos\theta-3\cos^2\theta-4\cos^2\delta\Big].
\eea
where $m^2\equiv (A)/(N_{A})$, $A$ being the background value of the $\vec{A}_{a}$'s evaluated at horizon crossing. \\Fig.2 shows a plot of the 'non-Abelian' bispectrum normalized to the ratio $(g_{c}^{2}H^2 m^2N_{A}^{4})/(k^{6}_{1}x_{2}^{2}x_{3}^{2})$, as a function of the angles $\theta$ and $\delta$ and for fixed values of $x^{*}$, $x_{2}$ and $x_{3}$

\bea\label{graph}\fl
B_{\zeta}(\theta,\delta)&\simeq &g_{c}^{2}\frac{H^2}{k_{1}^{6}}\frac{m^2 N_{A}^{4}}{10^{-1}}\Big[\cos^2\delta (8\cos\theta-1.4\times 10^3)+3\cos\delta(\cos^3\theta-\cos^2\theta-11)\nonumber\\\fl&-&11\cos 2 \delta-40-6\cos 2\theta-\cos\theta(3\cos^2\theta-30\cos\theta-10)\Big]
\eea
where we set $x^{*}=1$, while $x_{2}$ and $x_{3}$ were chosen in the 'squeezed' region, $x_{2}=0.9$ and $x_{3}=0.1$.

\end{document}